\documentclass[]{article}    
\usepackage{graphicx}
\usepackage{mediabb}
  
\date{2000 October 15}  
\begin{document}       

\title{\bf Rotation Curves of Spiral Galaxies}

\author{{\bf Yoshiaki SOFUE$^1$ and Vera RUBIN$^2$} \\
\\
{\it 1. Institute of Astronomy, University of Tokyo, Tokyo 181-0015, Japan }\\
{\it 2. Department of Terrestrial Magnetism, Carnegie Institution of Wash.}\\
{\it 5241 Broad Branch Road, N.W., Washington, DC 20015, USA}\\ 
\\
\\
{\it Appeared in Ann. Rev. Astron. \& Astrophys. Vol. 39, p.137, 2001}\\
\\ }
\maketitle

\def\v{\vskip 2mm}
\def\noi{\noindent}
\def\r{\hangindent 1pc  \noindent}
\def\cen{\centerline}
\def\endpage{\vfil\break}
\def\page{\vfil\break}
\def\kms{km s$^{-1}$}
\def\kmsmpc{km s$^{-1}$ Mpc$^{-1}$}
\def\vrot{$V_{\rm rot}$}
\def\Vrot{V_{\rm rot}}
\def\Msun{M_{\odot \hskip-5.2pt \bullet}}
\def\msun{$M_{\odot \hskip-5.2pt \bullet}$}
\def\halpha{H$\alpha$}
\def\Halpha{H$\alpha$}
\def\ha{H$\alpha$}
\def\Ha{H$\alpha$}
\def\ee{\end{equation}}
\def\be{\begin{equation}}
\def\deg{$^\circ$}

\begin{abstract}
Rotation curves of spiral galaxies are the major tool  for determining the
distribution of mass in spiral galaxies. They provide fundamental
information for understanding the dynamics, evolution
and formation of spiral galaxies.
We describe various methods to derive rotation curves, and
review the results obtained.
We discuss the basic characteristics of observed rotation curves
in relation to various galaxy properties, such as
Hubble type, structure, activity, and environment.
\end{abstract}

\section{HISTORICAL INTRODUCTION}

The rotation of galaxies was discovered in 1914, when Slipher (1914) detected
inclined absorption lines in the nuclear spectra of M31 and the
Sombrero galaxy, and Wolf (1914) detected inclined  lines in the
nuclear spectrum of M81.
This evidence led Pease (1918) to use the Mt. Wilson 60-inch
to $``$investigate the rotation of the great nebula in
Andromeda" by obtaining a minor axis long slit spectrum of
M31 with an exposure of 84 hours taken during clear
hours in August, September, and October, 1916, and a major axis spectrum taken
over 79 hours in August, September, and October, 1917.
The absorption lines extended only 1.5 arcminutes in radius along the
major axis, less than 2\% of the optical radius,
but were sufficient to show the steep nuclear velocity rise.
Later studies of M31's rotation by Babcock (1939) and Mayall (1951)
extended major axis rotation velocities to almost 2\deg\ from the nucleus,
but exposure times were tens of hours, and spectrographs had stability problems.
Interestingly, both Babcock's velocities for M31 and Humason's
unpublished velocities for NGC 3115  showed the last measured
point to have a rotation velocity of over 400 \kms\  (almost 2 times actual),
but consequently raised questions of mass distribution.

At the dedication of the McDonald Observatory in 1939, Oort's (1940)
comment that $``$...the distribution of mass [in NGC 3115] appears to bear
almost no relation to that of the light" seems from the view in 2000 to
have attracted little attention.
His conclusion concerning the mass distribution in NGC 3115 is
worth quoting, even 60 years later.  $``$In the outer parts of the
nebula the ratio {\it f}
of mass density to light density is found to be very high; and this
conclusion holds for whatever dynamical model we consider.
The spectrum of the nebula shows the characteristics of G-type dwarfs.
Since {\it f} cannot be much larger than 1 for such stars,
they can account for roughly only 1/2 percent of the mass; the remainder
must consist either of extremely faint dwarfs having an average ratio of
mass to light of about 200 to 1 or else of interstellar gas and dust".
From  a reanalysis of the (scattered) velocities for M31,
Schwarzschild (1954) concluded that the approximately flat rotation
curve was $``$not discordant with the assumption of equal mass and
light distribution."

The modern era of optical observations
of rotation velocities within spiral galaxies dates from Page's (1952)
and especially Margaret and Geoffrey Burbidge's (1960)
observations which exploited the new red sensitivity of
photographic plates to observe the H$\alpha$ and [NII] emission
lines arising from HII regions within spiral disks. Within a decade,
rotation curves existed for a few dozen galaxies,
most of them extending only over the initial velocity rise and the turnover of the velocities.

Early radio observations of neutral hydrogen in external galaxies showed a
slowly falling rotation curve for M31 (van de Hulst et al. 1957) and a
flat rotation curve for
M33 (Volders 1959). The first published velocity field ($``$spider diagram') was of
M31 (Argyle 1965). For M33, the flatness could be
attributed to the side lobes of the beam, and was consequently ignored.
Louise Volders must also have realized that a flat rotation curve
conflicted with the value of the Oort constants for our Galaxy,
which implied a falling rotation curve at the position of the sun.
Jan Oort was one of her thesis professors.
By the 1970s, flat rotation curves were routinely detected
(Rogstad and Shostak 1972), but worries about side bands still persisted,
and a variation in M/L across the disk was a possible explanation
(Roberts and  Rots 1973).

Surveys of galaxy observations from these early years by de
Vaucouleurs (1959) and of galactic dynamics by Lindblad (1959)
reveal the development of the observations and the interpretation
of the spiral kinematics.  They are historically notable because
they contains early references, many of which have faded into oblivion.
More recent (but still early) reviews include
de Vaucouleurs \& Freeman (1973),
 Burbidge \& Burbidge (1975), van der Kruit and Allen (1978).

Rotation curves are tools for several purposes:
for studying the kinematics of galaxies;
for inferring the evolutionary  histories and the role that interactions
have played;
for relating departures from the expected rotation curve Keplerian form
to the amount and distribution of dark matter;
for observing evolution by comparing rotation curves in distant
galaxies with galaxies nearby.
Rotation curves derived from emission lines
such as H$\alpha$, HI and CO lines are particularly useful to
derive the mass distribution in disk galaxies,
because they manifest the motion of interstellar gases of population I,
which have much smaller velocity dispersion, of the order of
$5 - 10 $ \kms, compared to the rotation velocities.
This allows us to neglect the pressure term in the equation of motion
for calculating the mass distribution in a sufficiently accurate
approximation.

Here, we review the general characteristics of rotation curves for spiral
galaxies as kinematic tracers, in relation to galaxy properties, such as
Hubble types, activity, structure, and environment.
These parameters are fundamental input for understanding the dynamics and
mass distribution, evolution, and formation of spiral galaxies.
Methods for analysis are described.
In general, the discussion is restricted to studies since 1980.
Higher-order, non-axisymmetric velocity components due to
spiral arms and bars are not emphasized here.
Although rotation curves of spiral galaxies are a major
tool for determining the distribution of mass in spiral galaxies,
we stress the observations rather than the mass determinations
or the deconvolutions into luminous and dark matter.

Numerous discussions of rotation properties are included in the
conference proceedings,  {\it Galaxy Dynamics} (Merritt et al. 1999),
{\it Dynamics of Galaxies} (Combes et al. 2000),
{\it Galaxy Disks and Disk Galaxies} (Funes \& Corsini 2001).
Reviews of dark matter as deduced from galaxy rotation curves can be
found in Trimble (1987), Ashman (1992), and
Persic \& Salucci (1997, and papers therein).

Spheroidal galaxies have been reviewed earlier
(Faber \& Gallagher 1979; Binney 1982; de Zeeuw \& Franx 1991).
Generally, measures of velocity and velocity dispersions are necessary  for
mass determinations in early type galaxies, although methods which we
describe below are applicable to the cold disks often found in the cores  of ellipticals, in extended disks of low-luminosity ellipticals (Rix et al. 1999),
and  in S0 galaxies.

Data for several million galaxies are available from huge databases
accessible on the World Wide Web.  Hypercat (Lyon/Meudon Extragalactic
Database http://www-obs.univ-lyon1.fr/hypercat/) classifies references to
spatially resolved kinematics (radio/optical/1-dimension/2-dim/velocity
dispersion, and more) for 2724 of its over 1 million galaxies.
NED (NASA/IPAC Extragalactic Database http://nedwww.ipac.caltech.edu/index.html)
contains velocities for 144,000 galaxies. The number of measured velocity
points is tabulated for each galaxy reference. Both of these sites contain
extensive literature references for galaxy data.
High spatial (HST) STIS spectroscopy preprints are found
in http://STScI.edu/science/preprints.

\section{THE DATA}

When Margaret and Geoffrey Burbidge (1960) initiated their observational
program to determine the kinematics and hence masses of spiral galaxies,
they were reproducing the technique employed by Pease (1918),
but with improvements.
Telescopes were larger, spectrographs were faster, photographic plates
were sensitive in the red. The strong emission lines of H$\alpha$ and [NII]
could be more easily detected and measured than the weak broad H and K
absorption lines.
Since the 1980s, larger telescopes  and improved detectors have existed
for optical, radio, and mm observations. The combination of
high spatial and high spectral resolution digital detectors and speedy
computers has permitted a sophistication in the velocity analyses (Section 3)
that will surely continue.

\subsection{\ha\ and Optical Measurements}

Optical astronomers have available several observing techniques for
determining rotation curves and velocity fields for both the ionized
gas and stars.  Traditional long slit spectra are still valuable for
deducing the rotation curve of a galaxy from emission lines
(Rubin et al. 1980, 1982, 1985; Mathewson et al. 1992, 1996;
Amram et al.  1992, 1994; Corradi et al. 1991; Courteau 1997;
Vega Baltran 1999), but methods which return
the entire velocity field, such as Fabry-Perot spectrographs (Vaughan 1989)
or integral (fiber-optic) field instruments (Krabbe et al. 1997) offer more
velocity information at the price of more complex and time-consuming reductions.
Although H$\alpha$, [NII], and [SII] emission lines have traditionally
been employed, the Seyfert galaxy NGC 1068 has become the first galaxy whose
velocity field has been studied from the IR [Si VI] line (Tecza et al. 2000).
Distant planetary nebula (Section 2.5) and satellite galaxies  are
valuable test particles for determining the mass distribution at large
distances from galaxy nuclei. For a limited number of nearby galaxies,
rotation curves can be produced from velocities of individual HII regions
in galactic disks (Rubin \& Ford, 1970, 1983;
Zaritsky et al. 1989, 1990, 1994).

\subsection{HI line}

The HI line is a powerful tool to obtain kinematics of spiral galaxies,
in part because its radial extent is often greater, sometimes 3 or 4
times greater, than that of the visible disk. Bosma's thesis (1981a, b;
van der Kruit \& Allen 1978)  played a fundamental role in establishing
the flatness of spiral rotation curves. Instrumental improvements in the last
20 years have increased the spatial resolution of the beam, so that problems
 arising from low resolution are important only near the nucleus or in special
  cases (Section 4). While comparison of the inner velocity rise for
NGC 3198 showed good agreement between the 21-cm and the optical velocities
(van Albada et al. 1985; Hunter et al. 1986), the agreement was poor for Virgo
 spirals observed at low HI resolution (Guhathakurta et al. 1988;
 Rubin et al. 1989).
For low surface brightness galaxies, there is still discussion over whether
the slow velocity rise is an attribute of the galaxy or due the
instrumentation and reduction procedures (Swatters 1999, 2001; de Blok et al
2001; Section 7.5).

\subsection{CO Line}

The rotational transition lines of carbon monoxide (CO) in the millimeter wave
range [e.g., 115.27 GHz for $^{12}{\rm CO} (J=1-0)$ line, 230.5 GHz
for $J=2-1$] are valuable in studying rotation
kinematics of the inner disk and central regions of spiral galaxies,
for  extinction in the central dusty
disks is negligible at CO wavelengths (Sofue 1996, 1997).
Edge-on and high-inclination galaxies are particularly useful for
rotation curve analysis in order to minimize the uncertainty arising
from inclination corrections, for which extinction-free measurements are
crucial, especially for central rotation curves.

Because the central few kpc of the disk are dominated by molecular gas
(Young \& Scoville 1992; Young et al. 1995; Kenny \& Young 1988;
Garcia-Burillo et al. 1993; Nakai et al. 1995; Nishiyama \& Nakai 1998;
Sakamoto 1999),
the molecular fraction,  the ratio of the
molecular-gas mass density to that of total of molecular and HI masses,
usually exceeds 90\% (Sofue et al. 1995; Honma et al. 1995).
CO lines are emitted from molecular clouds associated with
star formation regions emitting the \ha\ line.
Hence, CO is a good alternative to \ha\,  and also to HI in the
inner disk, while HI is often weak or absent in the central regions.
The \ha, CO, and HI rotation curves agree well with each other
in the intermediate region disks of spiral galaxies
(Sofue 1996; Sofue et al. 1999a, b).
Small displacements between \ha\ and CO rotation curves can arise in the
inner regions from the extinction of the optical lines and the contamination
of the continuum star light from central bulges.

Decades ago, single dish observations in the mm wave range had angular
resolutions limited from  several to tens of arc seconds
due to the aperture diffraction limit.
Recently, however, interferometric observations have achieved sub- or
one-arcsec resolution  (Sargent and Welch 1993; Scoville et al. 1993;
Schinnerer et al. 2000; Sofue et al. 2000), comparable to, or sometimes
higher than, the current optical measurements (Fig. 1).
Another advantage of CO spectroscopy is its high velocity
resolution of one to several \kms.

\begin{figure}
\begin{center}      
\includegraphics[width=7cm]{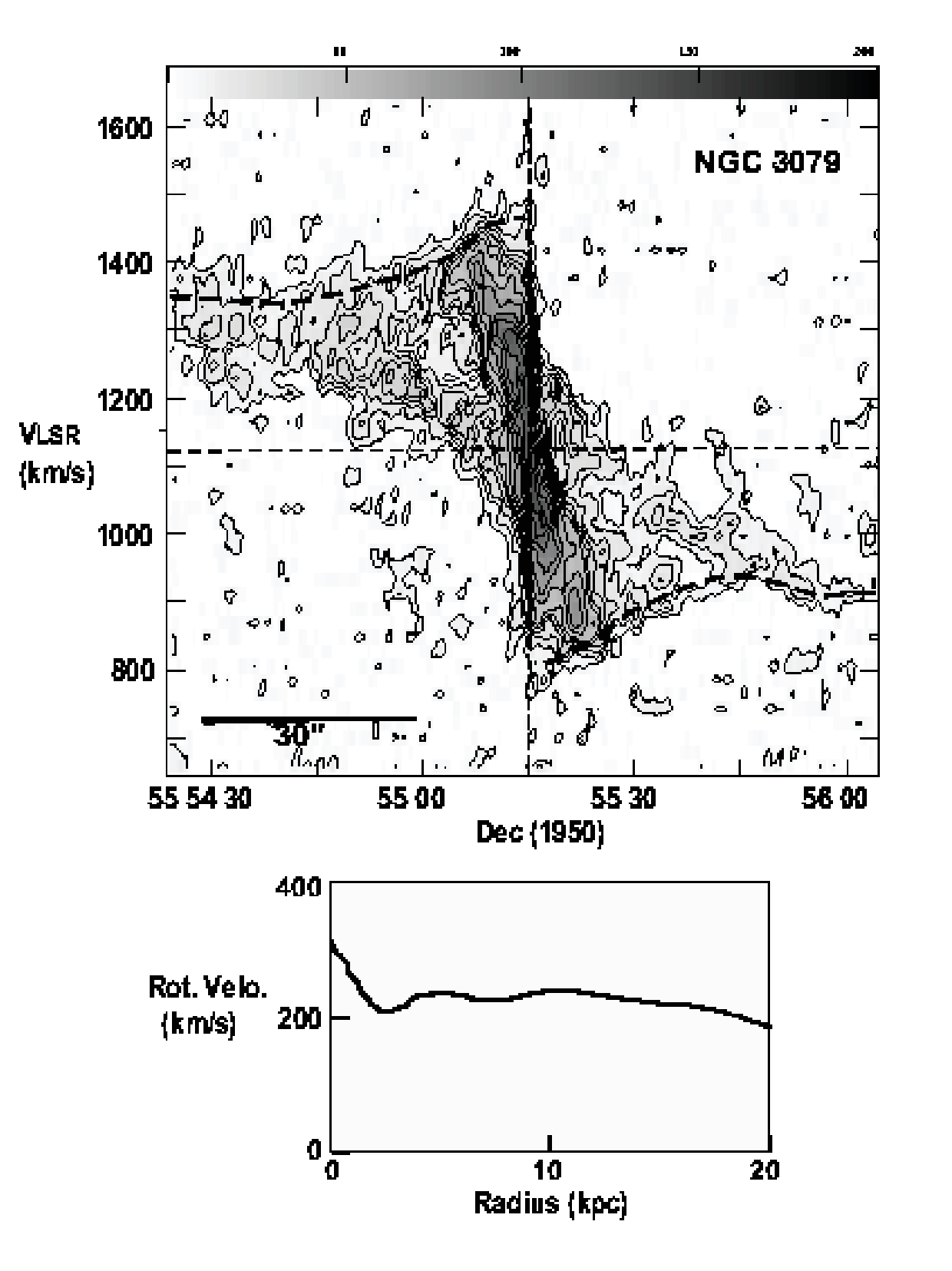}  
\caption{Position-velocity diagram along the major axis of the
edge-on galaxy NGC 3079 in the CO ($J=1-0$) line emission
at a resolution of $1''.5$ observed with the 7-element interferometer
consisted of the 6-element mm-wave Array and the 45-m telescope at Nobeyama
(Sofue et al. 2000).
The lower panel shows a composite rotation curve produced by combining the
CO result and HI data (Irwin and Seaquist 1991) for the outer regions.}
\end{center}
\label{Fig1}
\end{figure} 

\subsection{Maser Lines}

Radial velocity observations of maser lines, such as SiO, OH and H$_2$O
lines, from circum-stellar shells and gas clouds allow us to measure
the kinematics of stellar components in the disk and bulge of our Galaxy
(Lindqvist et al. 1992a, b; Izumiura 1995, 1999; Deguchi et al. 2000).
VLBI astrometry of SiO maser stars' proper motion and parallax
as well as radial velocities will reveal more unambiguous rotation of
the Galaxy in the future.
VLBI measurements of water masers from nuclei of galaxies reveal circumnuclear
rotation on scales of 0.1 pc around massive central black holes, as was
successfully observed for NGC 4258 (Miyoshi et al. 1995; see Section 4.4).

\subsection{Planetary Nebulae, Fabry-Perot, and Integral Field Spectrometers}

Planetary nebulae (PN) are valuable tracers of the velocity fields of early
type and complex galaxies, at large nuclear distances where the optical
light is faint or absent (Jacoby et al. 1990;
Arnaboldi  1998; Gerssen 2000),
and for galaxies in clusters (Cen A, Hui et al. 1995;
Fornax A, Arnaboldi et al. 1998). The simultaneous analysis of absorption
 line velocities for inner regions and hundreds of PN in the outer regions
 can constrain the viewing geometry as no single tracer can, and thus
reveal valuable details of the kinematics and the mass distribution
(Rix et al. 1997; Arnaboldi et al. 1998).

Fabry-Perot spectrometers are routinely used to derive the H$\alpha$ velocity
fields of spirals of special interest  (Vaughan 1989; Vogel et al. 1993;  Regan
and Vogel 1994; Weiner \& Williams 1996). Like Integral Field Spectrometry
(Krabbe et al. 1997), these techniques will
acquire more adherents as the instrument use and the reduction techniques
become routine.

\section{MEASURING ROTATION VELOCITIES}

Although the mathematics of rotating disks is well established
(e.g., Plummer 1911, Mestel, 1963, Toomre 1982, Binney \& Tremaine 1987,
Binney \& Merrifield, 1998) the analysis of the observational data has
continued to evolve as the quality of the data has improved.
Both emission lines
and absorption lines at a point on a spectrum are an integral along the line
of sight through the galaxy. Only  recently has the quality of the
observational material permitted the deconvolution of various components.
We describe a few procedures below.

\subsection{Intensity-Weighted-Velocity Method}

A rotation curve of a galaxy is defined as the trace of velocities on a
position-velocity (PV) diagram along the major axis, corrected for the
angle between the line-of-sight and the galaxy disk.
A widely used method is to trace intensity-weighted velocities
(Warner et al. 1973),
which are  defined by
\be
 V_{\rm int}=\int I(v) v dv / \int I(v) dv,
\ee
where $I(v)$ is the intensity profile at a given radius
as a function of the radial velocity.
Rotation velocity is then given by
\be
\Vrot=(V_{\rm int}- V_{\rm sys})/{\rm sin}~i,
\ee
where $i$ is the inclination angle and
$V_{\rm sys}$ is the systemic velocity of the galaxy.

\subsection{Centroid-Velocity and Peak-Intensity-Velocity Methods}

In outer galactic disks, where line profiles can be assumed to be
symmetric about the peak-intensity value, the intensity-weighted velocity can
be approximated by a centroid velocity of half-maximum values of a line
profile (Rubin et al. 1980, 1982, 1985),
or alternatively by a velocity at which the intensity attains
its maximum, the peak-intensity velocity (Mathewson et al. 1992, 1996).
Both methods have been adopted in deriving emission line rotation curves.
Tests indicate that centroid measures of weak emission lines show less scatter
(Rubin, unpublished).

However, for inner regions, where the line profiles are not
simple, but are superposition of outer and inner disk components,
these two methods often under-estimate the true rotation velocity.
The same situation occurs for edge-on galaxies, where line profiles are the
superposition of profiles arising from all radial distances sampled along
the line-of-sight. In these circumstances,
the envelope-tracing method described below
gives more reliable rotation curves.

\subsection{Envelope-Tracing Method}

This method makes use of the terminal velocity in a PV diagram along the
major axis.
The rotation velocity is derived by using the terminal velocity
$V_{\rm t}$:
\be
\Vrot=(V_{\rm t}-V_{\rm sys}) / {\rm sin}~i~
-(\sigma_{\rm obs}^2 + \sigma_{\rm ISM}^2)^{1/2},
\ee
where $\sigma_{\rm ISM}$ and $\sigma_{\rm obs}$ are
the velocity dispersion of the interstellar gas and the velocity resolution
of observations, respectively.
The interstellar velocity dispersion is of the order of
$\sigma_{\rm ISM} \sim 7$ to 10 \kms, while $\sigma_{\rm obs}$ depends on
instruments.

Here, the terminal velocity is defined by a velocity at which
the intensity becomes equal to
\be
I_{\rm t}=[(\eta I_{\rm max})^2+I_{\rm lc}^2]^{1/2}
\ee
on observed PV diagrams, where $I_{\rm max}$ and $I_{\rm lc}$ are
the maximum intensity and intensity corresponding to the lowest contour level,
respectively, and $\eta$ is usually taken to be 0.2 to 0.5.
For $\eta=0.2$,
this equation defines a 20\% level of the intensity profile at a fixed
position, $I_{\rm t}\simeq 0.2 \times I_{\rm max}$,
if the signal-to-noise ratio is sufficiently high. If the intensity is weak,
the equation gives
$I_{\rm t}\simeq I_{\rm lc}$ which approximately defines the loci along
the lowest contour level (usually $\sim 3 \times$ rms noise).

For nearly face-on galaxies observed at
sufficiently high angular resolution,
these three methods give an almost identical rotation curve.
However, both finite beam width and disk thickness
along the line of sight cause confusion of gas with
smaller velocities than the terminal velocity,
which often results in a lower rotation velocity in the former two methods.

The  envelope-tracing method is ill-defined when applied to
the innermost part of a PV diagram, for the two sides of the nucleus
have a discontinuity at the nucleus due
principally to the instrumental resolution, which is large with respect to
the velocity gradients. In practice, this discontinuity is avoided by
stopping the tracing
at a radius corresponding to the telescope resolution, and then
approximating the rotation curve by a straight
line crossing the nucleus at zero velocity. The $``$solid body" rotation
implied by this procedure is probably a poor approximation to the true
motions near the nucleus (Section 4.3).

\subsection{Iteration Method}

Takamiya and Sofue (private communication)
have developed an iterative method to derive a
rotation curve. This extremely reliable method comprises the following
procedure. An initial rotation curve, RC0, is adopted from a PV diagram (PV0),
obtained by any method as above (e.g. a peak-intensity method).
Using this rotation curve and an observed radial distribution of
intensity (emissivity) of the line used in the analysis, a PV diagram, PV1, is
constructed. The difference between this calculated PV diagram and the
original PV0, e.g. the difference between peak-intensity velocities, is used to
correct the initial rotation curve to obtain a corrected rotation curve, RC1.
This RC is used to calculated another PV diagram PV2 using the observed
intensity distribution, and to obtain the next iterated rotation curve, RC2
by correcting for the difference between PV2 and PV0.
This iteration is repeated until PV$i$ and PV0 becomes identical, such that
the summation of root mean square of the differences between PV$i$ and
PV0 becomes minimum and stable.
RC$i$ is adopted as the most reliable rotation curve.

\subsection{Absorption Line Velocities}

For several decades, the Fourier quotient technique (Simkin 1974,
Sargent et al.  1977) or the correlation technique (Bender 1990,
Franx \& Illingworth 1988) were methods of choice for determining
rotation velocities within early-type galaxies.
Both procedures assume that the   stellar absorption lines formed by
the integration along the line of sight through the galaxy can be
fit by a Gaussian profile. However, recent instrumental improvements
confirm that even disk galaxies consist of multi-component kinematic
structures, so more sophisticated methods of analysis
are required to reveal velocity  details of the separate stellar components.

Various methods have been devised to account for the non-Gaussian
form of the line-of-sight velocity distribution. Line profiles
can be expanded into a truncated Gauss-Hermite series
(van der Marel \& Franx 1993) which measure the asymmetric deviations
(h{$_3$}) and the symmetric deviations (h{$_4$}) from Gaussian.
Alternatively, one can use the unresolved
Gaussian decomposition method (Kuijken and Merrifield 1993).
Other procedures to determine line profiles and their higher
order moments (e.g. Bender 1990, Rix \& White 1992, Gerhard 1993)
are in general agreement (Fisher 1997); differences arise
from signal-to-noise, resolution,
and template mismatch. Such procedures will define the future
state-of-the-art.

\subsection{Dependence on Observational Methods}

Disk galaxies are a complex combination of various structural components.
Observations from emission lines and absorption lines in the optical, mm, and
radio regions may not sample identical regions along the same line-of-sight.
Instruments sample at  different sensitivities with different wavelength and
spatial resolutions.
Results are a function of the techniques of observations and reductions.
A simple $``$rotation curve" is an approximation as a function of radius
to the full velocity field of a disk galaxy.
As such, it  can be obtained only by neglecting small scale
velocity variations, and by
averaging and smoothing  rotation velocities from both sides of the
galactic center. Because it is a simple, albeit
approximate, description of a spiral velocity field, it is likely to be
valuable even as more complex descriptions become available for many galaxies.

\section{CENTRAL ROTATION CURVES}

Centers of galaxies are still mysterious places.  For galaxies as close as the
Virgo cluster, 0.1$''$ subtends about 8pc.  Only in a few special cases can
the stellar or gas kinematics be inferred on such scales: for very few
galaxies can we   sample velocity fields on scales of tens of parsecs.
Generally, when astronomers
discuss circumnuclear rotation curves, they refer to velocities  measured
from spectra where a single resolution element encompasses a large fraction of
the radius on which the velocities vary. $``$High accuracy" and $``$high
resolution" mean high with respect the present state-of-the-art.

\subsection{High Resolution and Dynamic Range}

A simulation reveals the effects of the finite resolution on the observed PV
diagram, for a galaxy with assumed gas and mass distributions (Sofue 1999a).
Fig. 2 shows an assumed rotation curve for a galaxy containing
a central compact core, bulge, disk
and massive halo, each expressed by a Plummer potential.
In the observed PV diagram, however, the central steep rise and
the peaks due to the core and bulge are
hardly recognized.
Central rotation curves derived from observed
PV diagrams generally give {\it lower limits} to the rotation velocities.
In fact, this conclusion holds for virtually all procedures
which do not adequately account for the finite observed resolution.

\begin{figure}
\begin{center}      
\includegraphics[width=6cm]{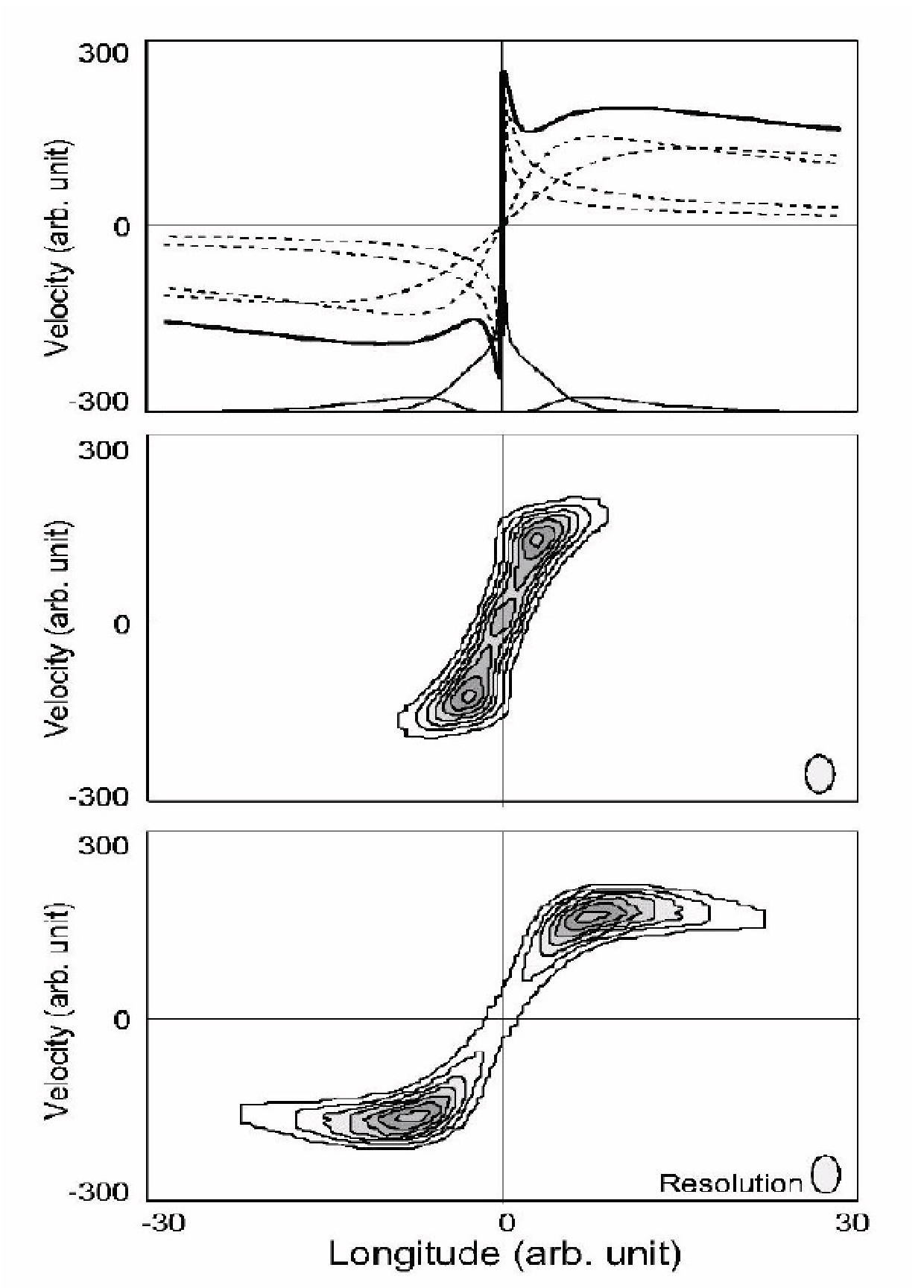}  
\caption{Simulation of the effect of beam-smearing on a position-velocity
diagram. The top panel shows an assumed 'true' rotation
curve comprising a central core,  bulge,  disk and a halo.
Assumed molecular and HI gas distributions are indicated by the thin lines.
The middle panel is an 'observed' PV diagram in CO, and the bottom in HI.
Both the resolution and sensitivity are crucial to detect central
high velocities and steep rise.}
\end{center}
\label{Fig2}
\end{figure} 

CCD spectroscopy has made it possible to derive optical rotation curves of
centers of galaxies, due to high dynamic range and
precise subtraction of bulge continuum light
(Rubin \& Graham 1987; Woods et al. 1990; Rubin et al. 1997;
Sofue et al. 1998, 1999a;
Bertola et al. 1998). However, optical spectroscopy often encounters additional
difficulty arising from
extinction due to dusty nuclear disks as well as confusion with absorption
features from Balmar wings of  A-type stars.
These problems are lessened at the wavelength of CO lines because of
the  negligible extinction, the high molecular gas content, and the high
spatial and velocity resolution. At present,
the combination of optical and CO-line spectroscopy produces rotation
curves of high accuracy, reliable for the entire regions of galaxies
including the central regions
(Sofue 1996, 1997, Sofue et al. 1997, 1998, 1999).

In a few special cases, nuclear disks have been studied with other techniques.
For several bright radio cores, HI absorption features
have revealed high-velocity central disks
(Ables et al. 1987; Irwin \& Seaquist 1991).
Rapidly rotating nuclear disks studied from their water maser emission
and very high resolution observations are discussed in Section 4.4.

\subsection{The Milky Way Center}

By its proximity, our Galaxy provides a unique opportunity
to derive a high resolution central rotation curve (Gilmore et al. 1990).
Proper-motion studies in the near infrared show
that the velocity dispersion of stars within the central 1 pc
increases toward the center, indicating the existence of a massive
black hole of mass $3\times 10^6\Msun$ (Genzel et al. 1997, 2000;
Ghez et al. 1998).

The rotation curve varies slightly depending upon the tracer.
A rotation curve formed from high resolution
CO and HI-line spectroscopy
(Burton \& Gordon 1978; Clemens 1985; Combes 1992),
shows a very steep rise in the central hundred
pc region, attaining a peak velocity of 250 \kms\ at $R\sim 300$ pc.
It then decreases to a minimum at $R\sim 3$ kpc of about 200 \kms,
followed by a gentle maximum at 6 kpc and a flat part beyond the solar
circle. Rotation velocities due to the black hole are combined
with the outer velocities in Fig. 3: the curve is presented
both in linear and logarithmic plots.
Of course, the rotation velocity does not decline to zero at the nucleus,
but increases inward, following a Keplerian law.

\begin{figure}
\begin{center}      
\includegraphics[width=7cm]{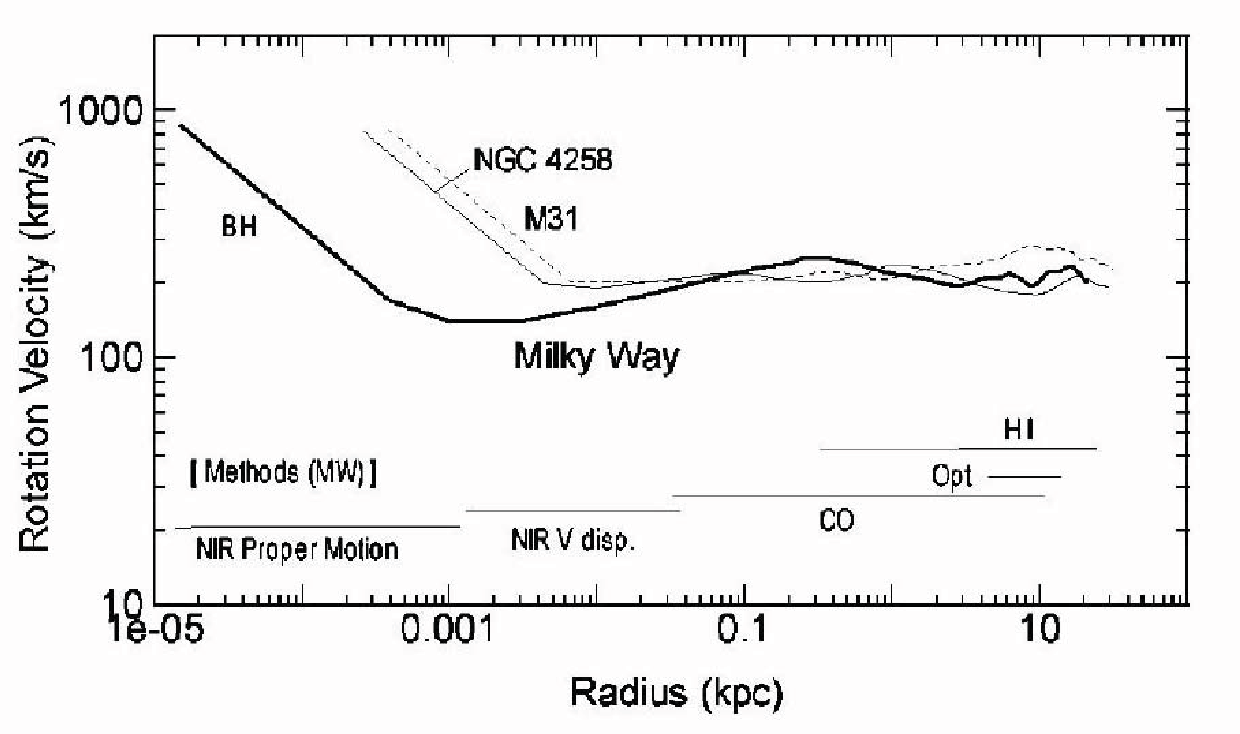}  
\caption{Logarithmic rotation curves of the Milky Way (thick line),
NGC 4258 (thin line) and M31 (dashed line).
Innermost rotation velocities are Keplerian velocities
calculated for massive black holes.
Observational methods for the Milky Way are shown by horizontal lines. }
\end{center}
\label{Fig3}
\end{figure} 

Radial velocities of OH and SiO maser lines from IR stars in the Galactic
Center region are used to derive the velocity dispersion and the mass
within the observed radius, as well as the mean rotation, which
seems to take part in the Galactic rotation
(Lindqvist et al.1992a, b; Sjouwerman et al. 1998).
SiO masers from IRAS sources in the central bulge
have been used to study the kinematics, and the mean rotation of the bulge
was found to be in solid body rotation of the order of 100 \kms
(Izumiura et al. 1995; Deguchi et al. 2000).
SiO masers in the disk region have been also used to study the structure
and kinematics of a possible bar structure and non-circular streaming motion
superposed on the disk and bulge components (Izumiura et al. 1999).

\subsection{Rapidly Rotating Central Components and Massive Cores}

Central rotation curves have been produced for a number of galaxies
by a systematic compilation of PV diagrams in the CO and \ha\ lines
(Sofue 1996; Sofue et al. 1997, 1998, 1999).
Fig. 4 shows  rotation curves obtained for nearby galaxies
at high  spatial and velocity resolution.
For massive spiral galaxies, high nuclear velocities may be a universal
property, but detected only with highest resolution observations.
Even a decade ago, it was observed (Rubin \& Graham 1987) that innermost
velocities  for some galaxies start from an already
high velocity at the nucleus. But high central density may not be a
characteristic only of massive galaxies. The nearby M33 (1$''$=3pc), a galaxy
with a minimal $``$bulge", exhibits velocities over the inner $\pm$200pc which
are flat at about V=100 \kms\
(Rubin 1987), and do not decrease to zero at the
origin.
Here too, the contribution from the falling density of a peaked central mass
exceeds the density contribution from the disk.

\begin{figure}
\begin{center}      
\includegraphics[width=7cm]{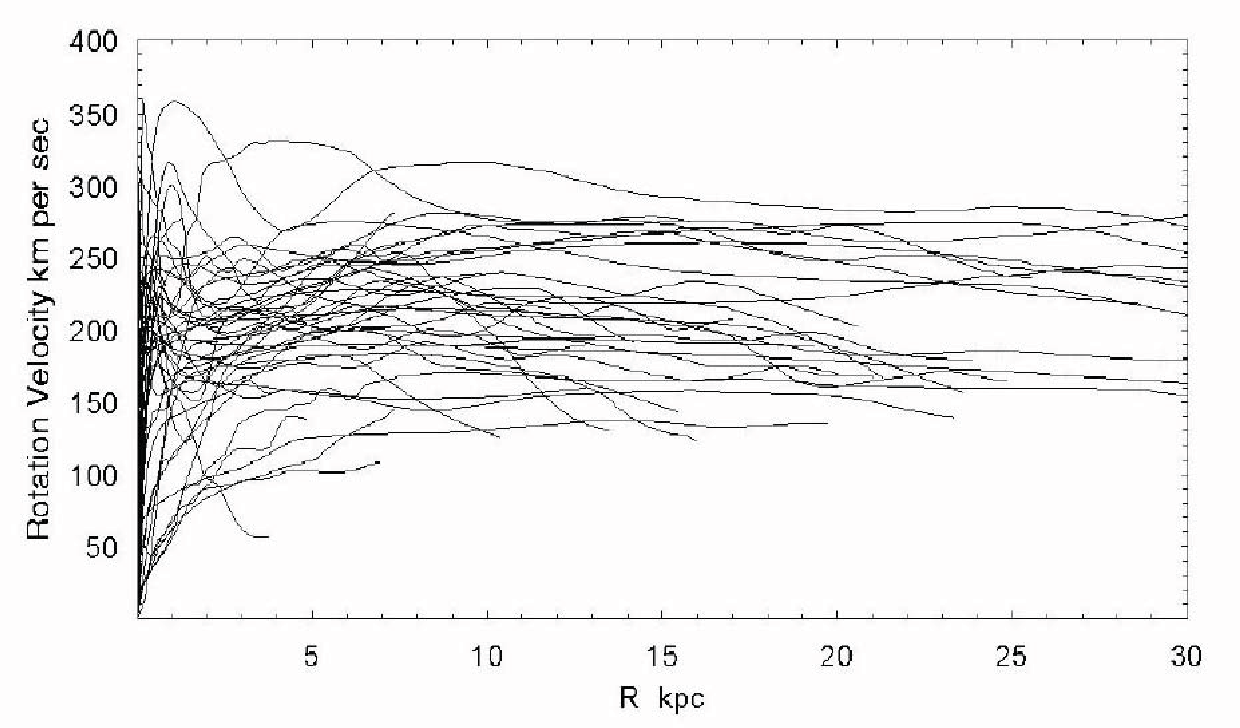}  
\caption{Rotation curves of spiral galaxies obtained by
combining CO data for the central regions, optical for disks,
and HI for outer disk and halo (Sofue et al. 1999). }
\end{center}
\label{Fig4}
\end{figure} 

Bertola et al. (1998) have emphasized that the
high-velocity nuclear peaks observed in some spiral
galaxies   match the simulated PV diagrams  for
Keplerian rotation due to a massive ($\sim 10^9\Msun$) black hole, at
equivalent resolution.  Even more dramatic, the analysis of
Maciejewski  \& Binney (2000) show that when a galaxy with an arbitrarily
large central velocity gradient is observed with a slit wider than the
instrumental point spread function, artifacts are generated in the spectra.
Such artifacts  can erroneously be interpreted as discrete kinematic
components, and may account for some of the features observed in the
spectra of Virgo galaxies (Rubin et al. 1999).

Evidence confirms that the steep nuclear rise observed in massive
galaxies is real, and not  due to a particular view of non-circular motions.
The probability of looking at a bar side-on
is larger than that of viewing one end-on. Hence there is
a larger probability for apparently slower rotation
than circular velocity.
For these massive galaxies, the mass density increases toward the
nucleus more rapidly than expected from an exponential or de Vaucouleurs law.
The widely adopted custom of drawing a rotation curve by linking
positive and negative velocities from the opposite sides across the
nucleus along the major axis is incorrect, at least for these massive galaxies.

\subsection{Massive Black Holes and Circum-nuclear Rotation}

For many spirals, the innermost region exhibits rapid rotation velocities
(Carter \& Jenkins 1993; van der Marel et al. 1994;
Miyoshi et al. 1995; Kormendy \& Richstone 1995; Richstone, et al. 1998,
Bertola et al. 1998; Ferrarese 1999; Kormendy \& Westpfahl 1989;
Kormendy 2001).
These high velocities offer evidence for massive nuclear black holes.
Consequently, orbital velocities in the center decrease rapidly from a
velocity close to the speed of light.
Detecting these high central velocities will require both
enormously high spatial and velocity resolution and is a program for
the future.

At present, spectroscopic sub arcsecond seeing is limited to the Space
Telescope Imaging Spectrographs(STIS) and
a few ground based telescopes, except in special cases.
STIS is now engaged in a major study of a sample of 54 Sb-Sc spirals,
V$\le$2000\kms\  to obtain spectra at H$\alpha$,
within a few arcsec of the nucleus (Axon, unpublished).
The aim is to measure black hole masses, or significant upper limits.

In one very special nearby galaxy, NGC 4258, water masers at 22 GHz
are observed  from a disk of radius 0.1 pc in
Keplerian rotation about a mass of $3.9 \times 10^{7}\Msun$ (Nakai et al. 1993;
Watson \& Wallim 1994; Miyoshi et al. 1995; Herrnstein et al. 1999).
The maximum rotation velocity is 900 \kms; the rotation period is 800 years.
VLBI observations of the water maser line
have revealed a rapidly rotating nuclear torus of sub parsec scales
in several nearby active galactic nuclei
(NGC 3079: Haschick et al. 1990; Trotter et al. 1998; Sawada-Satoh et al. 2000;
NGC 1068: Greenhill et al. 1996; NGC 4945: Greenhill et al. 1997).

\subsection{Activity and Rotation Curves}

One might ask whether the existence of massive objects in the nuclei,
as suggested from the high central velocities, is correlated with nuclear
activity.
High-accuracy central rotation curves for starburst galaxies
(NGC 253, NGC 1808, NGC 3034),
Seyferts (NGC 1068, NGC 1097), LINERs (NGC 3521, NGC 4569, NGC 7331),
and galaxies with nuclear jets (NGC 3079) (Sofue et al. 1999a;
Brinks et al. 1997) show, however, no particular peculiarity.
Even such a very active galaxy like NGC 5128 (Cen  A) shows a rotation
curve much like a normal galaxy (van Gorkom et al. 1990).
The radio lobe galaxy NGC 3079 has both strong nuclear activity
and usual rotation properties, but with very high central velocities
(Sofue \& Irwin 1992; Irwin \& Sofue 1992; Sofue et al. 1999a).
While these galaxies all show a very steep central velocity rise, such steep
rise is generally observed
for massive galaxies without pronounced central activity. Because
the global rotation and mass distribution in active spirals
are generally normal, it is likely that nuclear activity is triggered by local
and temporal causes around central massive cores and/or black holes.

\subsection{Resonance Rings}

The ring resonance in rotating disks will affect the kinematics
and rotation of gas and stars in a galaxy disk.
Rotation curves for several ring galaxies  (Buta et al. 1995, 1999) exhibit
normal rotation properties, showing a steep nuclear rise,  high-velocity
peak near the resonance ring, and flat velocities in the disk and halo.
Simulations of rotation properties for a bar resonance mimic well the
observed variations of rotation velocities, which is of the
order of $\pm 20 - 30$ \kms\ (Salo et al. 1999).

\subsection{ Nuclear Warp}

A major interest in current interferometer observations of the
CO line emissions from nuclear regions is the
detailed orientation of the nuclear molecular disk (NMD)
and circum-nuclear torus.
A NMD is produced by accretion of disk gas due to an
angular momentum transfer to the massive disk by galactic
shock waves, either in spiral arms or bars, whereas off-axis angular
momentum such as associated with warping is invariant.
If the accreting disk has a warp, as is often the case and is particularly
prominent in mergers and interacting galaxies,
the displacement of angular momentum of accreting disk from the original
rotation axis is amplified.
Hence, NMDs often exhibit significant warp from the main disk.
Interferomeric CO observations exhibit that
NGC 3079's NMD is warped from the main disk by 20 degrees
which contains a higher-density molecular core inclined from both the
NMD and main disk (Sofue et al. 2000);
NGC 1068 shows an warped nuclear disk surrounding a nuclear torus whose
axis is quite different from that of the main disk and nuclear disk
(Kaneko et al. 1997; Schinnerer et al. 2000).
A nuclear warp produces uncertain inclination corrections in the
rotation velocities. These can  be minimized by observing edge-on galaxies.

\subsection{Nuclear Counterrotation}

An extreme case of a nuclear warp is  counterrotation.
Rotating nuclear disks of cold gas have been discovered in  more than 100
galaxies, types E through Sc
(Bertola \& Galletta 1978; Galletta 1987, 1996; Bertola et al. 1990;
Bertola et al. 1992; Rubin 1994b; Garcia-Burillo et al. 1998);
counterrotation is not especially rare.
Simulations of disk interactions and mergers which  include gas and stellar
particles (Hernquist \& Barnes 1991; Barnes \& Hernquist 1992)
reveal that a kinematically distinct nuclear gas disk can form;
it may be counterrotating.
Simulation of galactic-shock accretion of nuclear gas disk
in an oval potential, such as a nuclear bar,
produces highly eccentric streaming motion toward the nucleus,
some portion being counterrotating (Wada et al. 1999).
Kinematically decoupled {\it stellar} nuclear disks are also observed in
early type galaxies (Jedrzejewski \& Schechter 1989;
Franx et al.  1991). Counter rotating nuclear disks can result from
merger, mass exchange and/or
inflow of intergalactic clouds.
In addition to forming the central disk,  an inflow of
counterrotating gas would also be likely to promote nuclear activity.

\subsection{Non-circular Motion in Nuclear Molecular Bar}

Oval potential such as due to a bar produces galactic shocks of
interstellar gas, and the shocked gas streams along the bar in
non-circular orbits (Sorensen et al. 1976;
Noguchi 1988; Wada \& Habe 1992, 1995; Shlosman 1990).
The velocity of streaming motion during its out-of-shock passage
is higher than the circular velocity, while the velocity
during its shock passage is much slower than circular velocity, close to
the pattern speed of the bar in rigid-body rotation.
The molecular gas is strongly condensed in the galactic shock, and stays
there for a large fraction of its orbiting period.
Hence, CO line velocities manifest the velocity of shocked gas,
and therefore, observed CO velocities are close to those of gas in
rigid-body motion with a bar, slower than the circular velocity.
This results in underestimated rotation velocities.
Geometrical effect that the probability of side-on view of a bar
is greater than that of end-on view also causes underestimated rotation
velocities.

Bar-driven non-circular motion of the order of 20 to 50 \kms\
are observed in central molecular disks
(Ishizuki et al. 1990; Handa et al. 1990;
Sakamoto et al. 1999; Kenney et al. 1992; Kohno 1998; Kohno et al. 1999).
The CO-line PV diagram in the Milky Way Center (Bally et al. 1985;
Oka et al. 1998; Sofue 1995) shows that the majority ($\sim 95$\%) of
gas is rotating in steep rigid-body features.
A few percent exhibits non-circular 'forbidden' velocities,
which could be due to non-circular motion in an
oval potential (Binney et al. 1991), whereas a question remains why
the majority of the gas is regularly rotating.
Because the gas is shocked and intensity-weighted velocities are smaller
than the circular velocity as discussed above, determination of mass
distribution in barred galaxies from observed CO velocities
is not straightforward, and will be a challenge for numerical
simulations in the future.


\section{DISK ROTATION CURVES}

A disk rotation curve manifests the  distribution of
surface mass density in the disk, attaining a broad maximum at a
radius of about twice the scale radius of the exponential disk.
For massive Sb galaxies, the rotation maximum appears at a radius
of 5 or 6 kpc, which is about twice the scale length of the disk.
Beyond the maximum, the rotation curve is usually flat,
merging with the flat portion due to the massive dark halo.
Superposed on the smooth rotation curve are fluctuations of
a few tens of \kms\ due to spiral arms or velocity ripples.
For barred spirals, the fluctuations are larger, of order 50 \kms,
arising from non circular motions in the oval potential.

\subsection{Statistical Properties of Rotation Curves}

The overall similarity of shapes of rotation curves for spiral galaxies
has led to a variety of attempts to categorize their forms, and to establish
their statistical properties.
Kyazumov (1984) cataloged rotation curve parameters for 116 normal
S and Ir galaxies, and categorized the shapes.
Rubin et al. (1985) formed families of Sa, Sb, and Sc synthetic rotation
curves as a function of luminosity, from the galaxies they had observed.
Casertano \& van Gorkom (1991), using HI velocities, studied
rotation curves as a function of luminosity.

Mathewson et al. (1992, 1996) used their massive set of \ha\
rotation curves together with optical luminosity profiles for
2447 southern galaxies,  to examine the Tully-Fisher (1977) relation.
For a subset of 1100 optical and radio rotation curves,
Persic et al. (1995, 1996) fit the curves by a formula, which is a
function of total luminosity and radius, comprising both disk and
halo components. Both the forms and amplitudes are functions of
the luminosity, and  the outer gradient of the RC is a decreasing
function of luminosity.  Their formula does not contain any free
parameters, and they call it universal rotation curve.
Courteau (1997) obtained optical long-slit rotation curves for 304
Sb-Sc northern UGC galaxies for Tully-Fisher applications, and
fitted the curves empirically by a simple function for the purpose to
calculate line widths.
Roscoe (1999) has attempted to parameterize outer-disk rotation
curves by an extremely simple power law of radius.

Universal rotation curves reveal the following characteristics.
Most luminous galaxies show  a slightly declining rotation curves in
the outer part, following a broad maximum in the disk.
Intermediate galaxies have nearly flat rotation from across the disk.
Less luminous galaxies have monotonically increasing
rotation velocities across the optical disk.
While Persic et al. conclude that the dark-to-luminous mass ratio
increases with decreasing luminosity, mass deconvolutions are far from unique.

A study of 30 spirals in the Ursa Major Cluster  (Verheijen 1997) showed
that 1/3 of the galaxies (chosen to have kinematically unperturbed gas disks)
have velocity curves which do not conform to the universal curve shape.
Like humans, rotation curves have their individualities, but they share many
common characteristics.  These common properties are meaningful in some
situations: in other circumstances their use may be misleading. It is
important to apply the common properties only in
appropriate situations, e.g., for outer disk and halo beyond $\sim 0.5$
optical radii, corresponding to several kpc for Sb and Sc galaxies.
Inner rotation curves have greater individuality (Sofue et al. 1999a).

\subsection{Environmental Effects in Clusters}

A variety of physical mechanisms can alter the internal kinematics of spirals
in clusters, just as these mechanism have altered the morphology of galaxies
in clusters (Dressler 1984; Cayatte et al. 1990).
Gas stripping, star stripping, galaxy-galaxy encounters, and interaction
with the general tidal field are all likely to occur. Early studies of optical
rotation curves for galaxies in clusters (Burstein et al. 1986;
Rubin et al. 1988; Whitmore et al. 1988, 1989)
detected a correlation between outer rotation-velocity gradients
and distances of galaxies from the cluster center. Inner cluster galaxies show
shallower rotation curves than outer cluster galaxies,  for distances
0.25 to 5 Mpc from cluster centers.
These authors suggest that the outer galaxy mass is truncated in the
cluster environment. Later studies have failed to confirm this result
(Amram et al. 1992, 1993, 1996; Sperandio et al. 1995).

A study of rotation curves for 81 galaxies in Virgo (Rubin et al. 1999,
Rubin \& Haltiwanger 2001) shows that about half (43) have rotation curves
identified as disturbed. Abnormalities include asymmetrical rotation
velocities on the two sides of the major axis, falling outer rotation curves,
inner velocity peculiarities, including velocities hovering near zero at
small radii, and dips in mid-disk rotation velocities.  Kinematic disturbance
is not correlated with morphology, luminosity, Hubble type, inclination,
maximum velocity, magnitude, or local galaxy density.

Virgo spirals with disturbed kinematics have a Gaussian distribution of
systemic velocities which matches that of the cluster ellipticals; spirals
with regular rotation show a flat distribution.  Both ellipticals and
kinematically disturbed spirals are apparently in the process of establishing
an equilibrium population. H$\alpha$ emission extends farther in the
disturbed spirals; the  gravitational interactions have also enhanced
star formation. The distribution on the sky and in systemic
velocity suggests that kinematically disturbed galaxies are on elongated
orbits which carry them into the cluster core, where galaxy-cluster and
galaxy-galaxy interactions are more common and stronger. Self-consistent
N-body models
that explore the first pass of two gravitationally interacting disk galaxies
(Barton et al. 1999) produce rotation curves with  mid-region velocity
dips matching those observed. Models of disk galaxies falling for the first
time into the cluster mean field (Valluri 1993) show m=1 (warp)
and m=2 (bar and spiral arms)
perturbations.

\subsection{Lopsided Position-Velocity Diagrams}

There is increased interest in galaxies with kinematically
lopsided HI profiles (Baldwin 1980; Sancisi 2001), which can arise from
a large-scale asymmetry of HI gas distribution in the spiral disk.
Of 1700 HI profiles, at least 50\% show asymmetries  (Richter \& Sancisi 1994);
recent work (Haynes et al. 1998) confirms this fraction. Because HI profiles
result from an integration of the velocity and the
HI distribution, {\it resolved} HI velocity fields offer more direct
information on kinematic lopsidedness.
From resolved HI velocity fields, Swaters et al. (1999) also estimate the
disturbed fraction to be at least 50\%.

As noted above, about 50\% of Virgo spirals show optical major
axis velocity disturbances; how this figure translates into lopsided
HI profiles is presently unclear.
The  field spirals studied earlier by Rubin et al. (1985),
chosen to be isolated and without obvious morphological peculiarities,
have rotation curves which are {\it very} normal (74\%; Rubin et al. 1999).
Yet the sample of optical rotation curves for galaxies in the
Hickson groups (Rubin et al. 1991) shows noticeably lopsided rotation curves
for $\ge$ 50\%. Further studies are needed to establish the frequency of
lopsidedness as a function of luminosity, morphology, HI content, resolution,
sensitivity, extent of the observations, and environment.

\subsection{Counterrotating Disks and Other Kinematic  Curiosities}

Only a handful of galaxies are presently known to have counterrotating
components over a large fraction of their disks (Rubin 1994b).
The disk of E7/S0 NGC 4550, (Rubin et al. 1992; Kenney \& Faundez 2000)
contains two hospital stellar populations, one
orbiting programmed, one retrograde.
This discovery prompted modification of  computer programs
which fit only a single
Gaussian to integrated absorption lines in galaxy spectra (Rix et al. 1992).
In NGC 7217 (Sab), 30\% of the disk stars orbit retrograde
(Merrifield \& Kuijen 1994).
The bulge in NGC 7331 (Sbc) may (Prada et al. 1996) or may not
(Mediavilla et al. 1998) counterrotate with respect to the disk.
Stars in NGC 4826 (Sab; the Black Eye or Sleeping Beauty) orbit
with a single sense.
Gas extending from the nucleus through the broad dusty lane rotates prograde,
but reverses its sense of rotation immediately beyond; radial infall
motions are present where the galaxy velocities reverse
(Rubin et al. 1965; Braun et al. 1994;  Rubin 1994a;
Walterbos et al. 1994; Rix et al. 1995; Sil'chenko 1996).

However, galaxies with extended counterrotating disks are not common.
A peculiar case is the early type spiral NGC 3593,
which exhibits two cold counterrotating disks (Bertola et al. 1996).
Of 28 S0 galaxies examined by Kuijken (1996), none have
counterrotating components accounting for more than 5\% of the disk light
(see also Kannappan \& Fabricant 2000).
Formation mechanisms for counterrotating disks can be devised
(Thaker \& Ryden 1998),  although cases of failure are reported
only anecdotally (Spergel, private communication).
While such galaxies are generally assumed to be remnants of mergers,
models show that generally the disk
will heat up and/or be destroyed in a merger.

In an effort to circumvent the problem of disk destruction in a merger,
Evans \& Collett (1994) devised a mechanism for producing
equal numbers of prograde and retrograde stellar disk orbits,
by scattering stars off a bar in a galaxy whose potential slowly changes
from triaxial to more axisymmetric.
In an even more dramatic solution by Tremaine \& Yu (2000;
see also Heisler et al. 1982; van Albada et al. 1982)
polar rings and/or counterrotating stellar disks can
arise in a disk galaxy with a triaxial halo. As the pattern speed of the
initially retrograde halo  changes to prograde due to infalling dark matter,
orbits of disk stars caught at the Binney resonance can evolve from
prograde to retrograde disk orbits. If, instead, the halo rotation
decays only to zero, stars with small inclinations are
levitated (Sridhar \& Touma 1996) into polar orbits.
This model predicts that stellar orbits in a polar ring will be divided
equally into two! counterrotating streams, making this a perfect
observing program for a very large telescope.

As spectral observations obtain higher resolution and sensitivity, emission
from weaker components is measured.
HI observations of  NGC 2403 reveal a normal
rotation, plus more slowly rotating HI extensions in a PV diagram,
the $``$beard", which can be due to an
infall of gas from an extended HI halo with slower rotation,
perhaps from a galactic fountain flow (Schaap et al. 2000).
Higher sensitivity VLA observations of NGC 2403 have
detected HI in the forbidden  velocity quadrants, which can be due
to a radial inflow (Fraternali et al. 2000).
Improved instrumentation permits detection of weaker features,
so we glimpse the kinematic complexities which exist in minor populations
of a single galaxy.

Edge-on  and face-on spirals are fine laboratories for studying
vertical kinematics.
In the edge on NGC 891, HI extends about 5 kpc above the plane,
where it rotates at about 25 \kms, more slowly than the disk
(Swaters et al. 1997).
Slower rotation is also observed in the CO halo in the edge on dwarf
M82 (Sofue et al. 1992).
Face on galaxies like M101 and NGC 628 show often extended
HI disk showing different kinematical properties from the
disk (Kamphuis et al. 1991; Kamphuis \& Brigg 1992).

\subsection{Rotation of High Redshift Galaxies}

Only recently have rotation curves been obtained for distant galaxies,
using HST and large-aperture ground-based telescopes with sub-arc second
seeing.
We directly observe galaxy evolution by studying galaxies closer to
their era of formation. Rotation velocities for moderately distant
spirals, z$\approx$ 0.2 to 0.4, (Bershady 1997, et al. 1999,
Simard \& Prichet 1998, Kelson et al. 2000a)  have already been surpassed with
Keck velocities reaching z$\approx$1 (Vogt et al. 1993, 1996, 1997; Koo 1999),
for galaxies whose diameters subtend only a few seconds of arc.
The rotation properties are similar to those of nearby galaxies,
with peak velocities between 100 to 200 \kms, and flat outer disk velocities.

Regularly rotating spiral disks existed at z$\approx$1, when the universe was
less than half of its present age. The Keck rotation velocities define a
TF relation (i.e., the correlation of rotation velocity
with blue magnitude) which matches to within $\le$0.5 magnitudes that for
nearby spirals.
Spiral galaxy evolution,  over the last half of the age of the universe,
has not dramatically altered the TF correlation.

\subsection{Rotation Velocity as a Fundamental Parameter of Galaxy Dynamics
and Evolution}

The maximum rotation velocity, reached
at a few galactic-disk scale radii for average and larger sized
spiral galaxies, is equivalent to one-half the velocity width of an integrated
21 cm  velocity profile.
The Tully-Fisher relation (1977; Aaronson et al. 1980; Aaronson \& Mould 1986),
the correlation between total velocity width and spiral absolute magnitude,
represents an oblique projection of the fundamental plane of spiral galaxies,
which defines a three-dimensional relation among the radius, rotation velocity,
and luminosity (absolute magnitude) (Steinmetz \& Navarro 1999;
Koda et al. 2000a,b).
The shape of a disk rotation curve manifests the mass distribution
in the exponential-disk, which is the result of dissipative process of
viscous accreting gas through the proto-galactic disk evolution
(Lin \& Pringle 1987).

As such, it emphasizes
the essential role that rotation curves play in determining the principal
galactic structures, and in our understanding of the formation
of disk galaxies (e.g. Mo et al. 1998).
For elliptical galaxies, the three parameter (half-light radius,
surface brightness, and central velocity dispersion) fundamental plane
relation (Bender et al. 1992; Burstein et al. 1997) is a
tool for studying elliptical galaxy evolution, analogous to the
spiral TF relation.
Keck spectra and HST images  of 53 galaxies in cluster CL1358+62  (z=0.33)
define a fundamental plane similar to that of nearby ellipticals
(Kelson et al. 2000b).
Ellipticals at z=0.33 are structurally mature; data for more distant
ellipticals should be available within several years.

\section{HALO ROTATION CURVES AND DARK MATTER: A Brief Mention}

The difference between the matter distribution implied by the
luminosity, and the distribution of mass implied by the rotation velocities,
offers strong evidence that spiral galaxies are embedded in extended
halos of dark matter. The physics of dark matter has been and will be
one of the major issues to be studied by elementary particle
physicists and astronomers.

\subsection{Flat Rotation Curve in the Halo}

When Rubin \& Ford (1970) published the rotation curve of M31,
formed from velocities of 67 HII regions, they noted that the mass
continued to rise
out to the last measured region, 24 kpc. They concluded  $``$extrapolation
beyond that distance is clearly a matter of taste".
By 1978, Rubin et al.  (1978) had learned  that $``$rotation curves of
high luminosity spiral galaxies are flat, at nuclear distances as
great as 50 kpc" ($H_0=50$ \kmsmpc).
Flat HI rotation curves were first noticed (Roberts \& Rots 1973) using a
single dish telescope.
However, it would be a few years before the observers and the theorists
(Ostriker \& Peebles 1973; Ostriker et al. 1974,  Einasto et al. 1974)
recognized each others' work, and collectively  asserted that disk
galaxies are immersed in extended dark matter halos.

Deeper and higher-resolution HI observations with synthesis telescopes
reveal that for the majority of spiral galaxies, rotation curves remain flat
beyond the optical disks (Bosma 1981a, b; Guhathakurta et al. 1988;
van Albada et al. 1985; Begeman 1989).
The Sc galaxy UGC 2885 has the largest known HI disk, with HI radius of
120 kpc for $H_0=50$ \kmsmpc\ (85 kpc
for $H_0=70$ \kmsmpc); the HI rotation curve is
still flat (Roelfsema \& Allen 1985).

The conclusion that a flat rotation curve is due to a massive dark halo
surrounding a spiral disk requires that Newtonian gravitational theory
holds over cosmological distances. Although proof of this assumption
is lacking, most astronomers and physicists prefer this explanation to
the alternative, that   Newtonian
dynamics need modification for use over great distances.
For readers interested in such alternatives,
see Milgrom (1983), Sanders (1996), McGaugh \& de Blok (1998),
de Blok \& McGaugh (1998), Begeman et al. (1991), Sanders (1996),
Sanders \& Verheijen (1998).
Non gravitational acceleration of halo gas rotation
would be also an alternative, such
as due to magneto-hydrodynamical force (Nelson 1998).

\subsection{Massive Dark Halo}

One of the best indicators of dark matter is the difference between the
galaxy mass predicted by the luminosity and the mass predicted by the
velocities.
This difference, which also produces a  radial variation of
the mass-to-luminosity ratio ($M/L$), is  a clue to the distribution of
visible and dark (invisible) mass (e.g., Bosma 1981a, b; Lequeux 1983;
Kent 1986, 1987;  Persic \& Salucci 1988, 1990;
Salucci \& Frenk 1989; Forbes 1992;  Persic et al. 1996;
H\'{e}raudeau \& Simien 1997; Takamiya \& Sofue 2000).
Unfortunately, there is not yet a model independent procedure for
determining the fraction of mass contained in the bulge, disk,
and dark halo, and mass deconvolutions are rarely unique.
Most current investigations
assume that the visible galaxy consists of a bulge and a disk,
each of constant $M/L$. Kent (1986, 1991, 1992) has used
the ``maximum-disk method'' to derive averaged $M/L$s in the
individual components.
Athanasoula et al. (1987) attempted to minimize the uncertainty
between maximum and minimal disks by introducing constraints to
allow for the existence of spiral structure.
Discussions of maximum or non-maximum disk persist
(Courteau and Rix 1999), even for our Galaxy (van der Kruit 2000).

Radial profiles of the surface-mass density (SMD) and surface
luminosity can be used to calculate $M/L$ directly.
Forbes (1992) derived the radial variation in the ratio of the
total mass to total luminosity involved within a radius,
$r$, an 'integrated $M/L$'.
Takamiya and Sofue (2000) determine
the SMD directly from the rotation curves, which can be sandwiched
by mass distributions calculated from rotation curves on both
spherical and flat-disk assumptions by solving  directly the formula
presented by Binney and Tremaine (1987).
A comparison of the SMD distributions  with optical surface
photometry shows that the radial distributions of the $M/L$ ratio
is highly variable within the optical disk and bulge, and
increases rapidly  beyond the disk,
where the dark mass dominates.

Separation of halo mass from disk mass, whether dark or luminous,
is an issue for more sophisticated observations and theoretical modeling.
Weiner et al. (2000a, b) have used non-circular streaming motion to
separate the two components using their theory that the streaming
motion in a bar potential is sensitive to the halo mass.

\subsection{The Extent of the Milky Way Halo}

Interior to the Sun's orbit, the mass of the Galaxy is
$\approx 10^{11}\Msun$.
Although there is evidence that the halo rotation curve is
declining beyond 17 kpc in a Keplerian fashion
(Honma \& Sofue 1997a), the mass distribution beyond the HI disk, e.g.
at $> 22$ kpc, is still controversial.
Interior to the distance of the Large Magellanic Cloud, the
Galaxy mass may grow to $6 \times 10^{11}\Msun$ (Wilkinson \& Evans 1999),
which depends upon the assumed orbit of the Cloud.
Interior to 200 kpc, the mass is at least $2 \times 10^{12}\Msun$
(Peebles 1995), matching masses for  a set of Milky Way-like galaxies
with masses inferred statistically from the velocities of their satellite
galaxies (Zaritsky  1992).

A Milky Way halo which extends at least 200 kpc is getting close to the
half-way distance between the Galaxy and M31, 350 kpc.
And if halos are as large as those suggested by the gravitational distortion
of background galaxies seen in the vicinity of foreground galaxies
(Fischer et al. 2000; Hoekstra 2000), then the halo of our Galaxy may brush
the equivalently large halo of M31.

\subsection{Declining Rotation Curves}

Few spirals exhibit a true Keplerian decline in their rotation velocities.
Among peculiar rotation curves,  declining rotation curves are occasionally
observed, confirming the conventional belief that the mass
distribution is truncated at about 1 to 3 optical radii (3-5 scale lengths)
(Casertano 1983; Casertano \& van Gorkom 1991; Barteldrees \& Dettman 1994).
Yet some galaxies exhibit Keplerian rotation curves well beyond the critical
truncation radius (Honma \& Sofue 1997a, b).
While truncation is an important issue for those who wish to weaken
the notion of a conspiracy of luminous and dark matter (Casertano
\& van Gorkom 1991), the issue is far from resolved.

\section{GALAXY TYPES AND ROTATION CHARACTERISTICS}

There is a marked similarity of form, but not of amplitude, of
disk and halo rotation curves
for galaxies with different morphologies from Sa to Sc (Rubin et al. 1985).
Thus the form of the gravitational potential in the disk and halo
is not strongly dependent on the form of the optical luminosity distribution.
Some moderate correlation is found between total luminosity
and rotation velocity amplitude.
Also, less luminous galaxies tend to show increasing outer rotation curve,
while most massive galaxies have slightly declining rotation in the
outmost part (Persic et al. 1996).
On the other hand, form of central rotation curves depend on the total
mass and galaxy types (Sofue et al. 1999):
Massive galaxies of Sa and Sb types show a steeper rise and
higher central velocities within a few hundred pc of the nucleus
compared to less massive Sc galaxies and dwarfs.
Dwarf galaxies generally show a gentle central rise.

\subsection{Sa, Sb, Sc Galaxies}

The maximum rotation velocities for Sa galaxies are higher than those of Sb
and Sc galaxies with equivalent optical luminosities.
Median values of $V_{\rm max}$ decreases from 300 to 220 to 175 \kms\ for
the Sa, Sb, and Sc types, respectively (Rubin et al. 1985).

Sb galaxies have  rotation curves with  slightly lower
values of the maximum velocity than  Sa (Rubin et al. 1982).
The steep central rise at 100-200 pc  is often associated
with a velocity peak at radii $r \sim 100-300$ pc
(Sofue et al. 1999a).
The rotation velocity then declines to a minimum at $r\sim 1 $ kpc,
and is followed by a gradual rise to a broad maximum at
$r \sim 2-7$ kpc, arising from the disk potential. The disk rotation curve has
superposed  amplitude
fluctuations of tens of \kms\ due to spiral arms or velocity ripples.
The outermost parts are usually flat, due to the massive dark halo.
Some Sb galaxies show a slight outer decline, often no larger than the inner
undulations  (Honma \& Sofue 1997a, b).

The rotation curve of the Milky Way Galaxy, a typical Sb galaxy,
is shown in Fig. 2 (Clemens 1985; Blitz 1979; Brand and Blitz 1993;
Fich et al. 1989; Merrifield 1992; Honma and Sofue 1995).
The rotation curve of Sb galaxies, including the Milky Way,
can be described as having:
(a) a high-density core, including the massive black hole, which causes a
non-zero velocity very close to the center;
(b) a steep rise within the central 100 pc;
(c) a maximum at radius of a few hundred pc, followed by a decline to a
minimum at 1 to 2 kpc; then,
(d) a gradual rise from to the disk maximum at 6 kpc; and
(e) a nearly flat outer rotation curve.

Sc galaxies have lower maximum velocities than Sa and Sb
(Rubin et al. 1980; 1985), ranging from $\le 100$ to $\sim 200$ \kms.
Massive Sc galaxies show a steep nuclear rise similar to Sb's.
However, less-massive Sc galaxies have a more gentle rise.
They also have a flat rotation to their outer edges.
Low-surface brightness Sc galaxies have a gentle central
rise with monotonically increasing rotation velocity toward the edge,
similar to dwarf galaxies (Bosma et al. 1988).

\subsection{Barred Galaxies}

Large-scale rotation properties of SBb and SBc galaxies are generally
similar to those of non-barred galaxies of Sb and Sc types.
However, the study of their kinematics is more complicated than for
non-barred spirals, because their gas tracers are less uniformly
distributed (Bosma 1981a,1996), and their iso-velocity contours are
skewed in the direction toward the bar (H$\alpha$, Peterson et al. 1978; HI,
Sancisi et al. 1979; stellar absorption lines, Kormendy 1983).
CO-line mapping and spectroscopy reveal high concentration of molecular
gas in shocked lanes along a bar superposed by significant non-circular
motions (Handa et al. 1990; Sakamoto et al. 1999; Kuno et al.  2000).
Thus, barred galaxies show velocity jumps from
$\pm \sim 30 - 40$ \kms\ to $\ge 100$ \kms\
on the leading edges of the bar, $R\sim 2 - 5$ kpc,
whereas some barred galaxies show flat rotation
(e.g., NGC 253: Sorai et al. 2000).
Non-barred spirals can show velocity variation
of about $\pm 10 - 20$ \kms, caused mainly by spiral arms.
Compared with non-barred spirals, barred  galaxies require a more
complete velocity field to understand their kinematics.
As discussed earlier, intensity-weighted velocities are underestimated
compared to the circular velocities, which is particularly crucial for shock
compressed molecular gas in the central regions (see Section 4.9).

This large velocity variation arises from the barred
potential of several kpc length.
Simulations of PV diagrams for edge-on barred galaxies
show many tens of \kms\ fluctuations,
superposed on the usual flat rotation curve
(Athanassoula and Bureau 1999; Bureau and Athanassoula 1999;
Weiner \& Sellwood 1999).
However, distinguishing the existence of a bar and quantifying it
are not uniquely done from such limited edge-on information.
For more quantitative results, two-dimensional velocity analyses are
necessary (Wozniak \& Pfenniger 1997).
In these models, barred spirals contain up to 30\% counterrotating stars;
the orbits are almost circular and perpendicular to the bar.
Pattern speeds for the bar have been determined from absorption line spectra
(Buta et al. 1996; Gerssen 2000 and references therein).

Due to their kinematic complexity, barred galaxies have been observed
considerably less than non-barred,
even though they constitute a considerable fraction of all disk galaxies
(Mulchaey \& Regan 1997).
However, high resolution optical observations, combined with HI and
CO,  have helped to stimulate the study of
two-dimensional non-circular velocity fields (e.g. Wozniak \& Pfenniger 1997,
Hunter \& Gottesman 1996; Buta et al. 1996). Gas streaming motions
along the bar are an efficient way to transport gas to the nuclear
regions (Sorensen et al. 1976;
Schwartz 1981; Noguchi 1988; Wada \& Habe 1992, 1995;
Wada et al. 1998; Shlosman et al. 1990), and lead to enhanced star formation.

\subsection{Low Surface Brightness Galaxies; Dwarf Galaxies}

Until the last decade, observations of rotational kinematics were
restricted to  spirals with average or high surface brightness.
Only within the past decade have low surface brightness (LSB)
galaxies been found in great numbers (Schombert \& Bothun 1988;
Schombert et al. 1992); many are spirals.
Their kinematics were first studied by de Blok et al. (1996) with HI,
who found slowly rising  curves which often
continued rising to their last measured point.
However, many of the galaxies are small in angular extent,
so observations are subject to beam smearing.
Recent optical rotation curves (Swaters 1999, 2001; Swaters et al. 2000;
de Blok et al. 2001) reveal a steeper rise for some,
but not all, of the galaxies studied previously at 21-cm.
It is not now clear if LSB galaxies are as dominated by dark matter
as they were previously thought to be; the mass models
have considerable uncertainties.

Dwarf galaxies, galaxies of low mass, are often grouped with low
surface brightness galaxies, either by design or by error.
The two classes overlap in the low surface brightness/low mass region.
However, some low surface brightness galaxies are large and massive;
some dwarf galaxies have high surface brightness.
Early observations showed dwarf galaxies to be slowly rotating,
with rotation curves which  rise monotonically to the last measured point
(Tully et al. 1978; Carignan \& Freeman 1985;
Carignan \& Puche 1990a,b; Carignan \& Beaulieu 1989;
Puche et al. 1990, 1991a, b; Lake et al. 1990; Broeils 1992).
The dark matter domination of the mass of the
dwarf galaxy NGC 3109 (Carignan 1985; Jobin \& Carignan 1990)
is confirmed by a reanalysis including H$\alpha$ Fabry-Perot
data (Blais-Ouellette et al. 2000a, b).
An exceptional case of a declining outer rotation curve has
been found in the dwarf galaxy NGC 7793 (Carignan \& Puche 1990a).
\ha\ velocity field observations of blue compact galaxies,
with velocities less than 100 \kms, show that
rotation curves rise monotonically to the edges of the galaxies
(\"Ostlin et al. 1999).

Swaters (1999) derived rotation curves from velocity fields obtained with the
Westerbork Synthesis Radio Telescope for 60 late-type dwarf galaxies of low
luminosity. By  an interactive analysis,  he obtained rotation curves which
are corrected for a large part of the beam smearing. Most of the rotation
curve shapes are similar to those of more luminous spirals; at the lowest
luminosities, there is more variation in shape.
Dwarfs with higher central light concentrations have more steeply
rising rotation curves, and a similar dependence is found
for disk rotation curves of spirals (Fig. 5).
For dwarf galaxies dominated
by dark matter, as for LSB (and also HSB) spirals, the contributions of the
stellar and dark matter components to the total mass cannot be unambiguously
derived. More high quality observations and less ambiguous mass
deconvolutions, perhaps more physics, will be required to settle questions
concerning the dark matter fraction as a function of mass and/or luminosity.

\begin{figure}
\begin{center}      
\includegraphics[width=7cm]{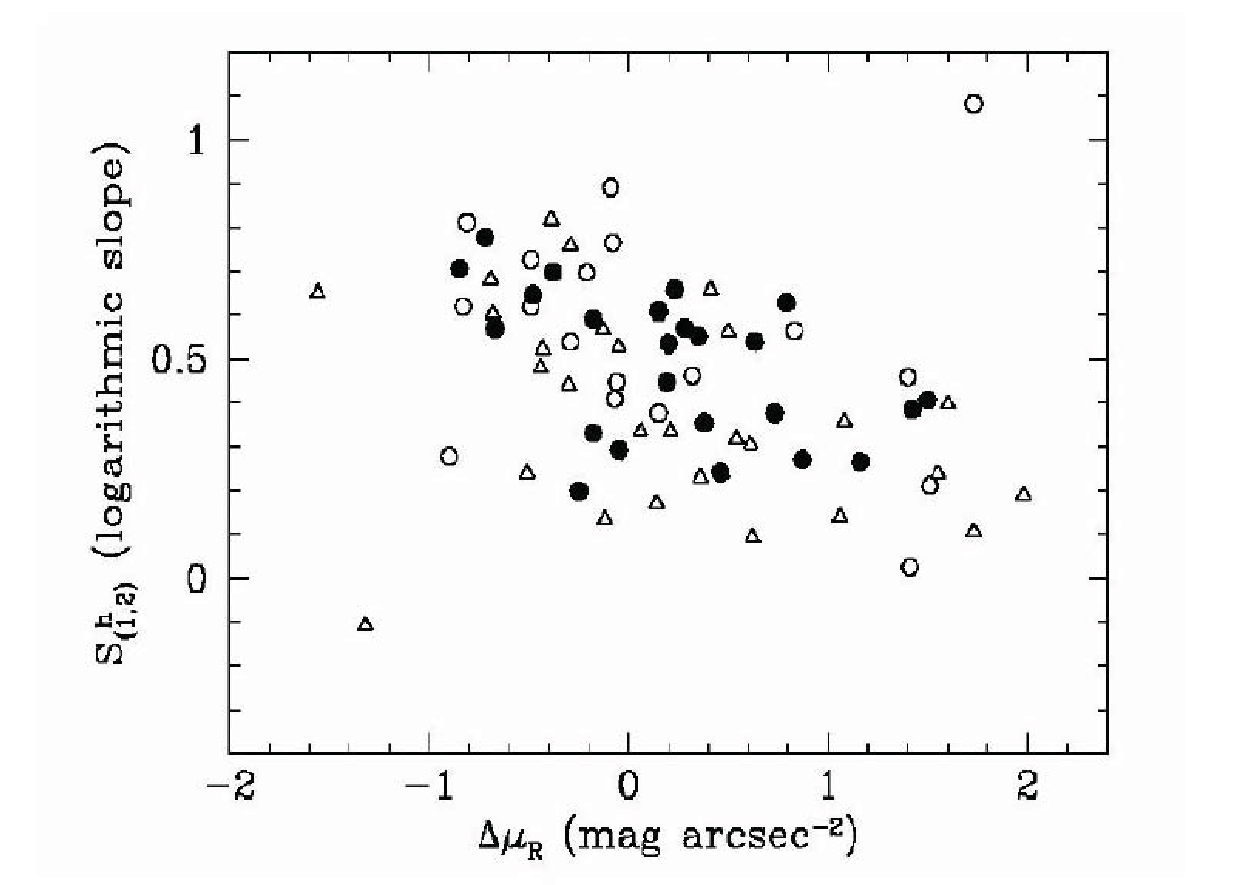}  
\caption{The logarithmic  slope of rotation curves between one and two
scale radii, plotted against the central concentration of light,
$\Delta\mu_{R}=\mu_0-\mu_{\rm c}$, where $\mu_{\rm c}$ is the
observed central surface brightness and $\mu_0$
the extrapolated surface brightness of the exponential disk.
Galaxies with larger excess of central brightness have flatter rotation.
Filled and open circles are dwarfs (high and low accuracy), and
triangles are spirals. [Courtesy of R. Swaters (1999)] }
\end{center}
\label{Fig5}
\end{figure} 

\subsection{Large Magellanic Cloud}

The LMC is a dwarf galaxy showing irregular optical morphology,
with the enormous starforming region, 30 Dor, located significantly
displaced from the optical bar and HI disk center.
High-resolution HI kinematics of the
Large Magellanic Cloud, Kim et al. (1998;
see Westerlund 1999 review)  indicate, however, a regular
rotation around the kinematical center, which is displaced
1.2 kpc from the center of the optical bar as well as from the center of
starforming activity (Fig. 6).
The rotation curve has a steep central rise, followed by a flat rotation
with a gradual rise toward the edge.
This implies that the LMC has a compact bulge
(but not visible on photographs),
an exponential disk, and a massive halo.
This dynamical bulge is 1.2 kpc away from the center of the stellar bar,
and is not associated with an optical counterpart.
The $``$dark bulge" has a large fraction of dark matter,
with an anomalously high mass-to-luminosity ($M/L$) ratio (Sofue 1999).
In contrast, the stellar bar has a smaller $M/L$ ratio compared to that
of the surrounding regions.

\begin{figure}
\begin{center}      
\includegraphics[width=7cm]{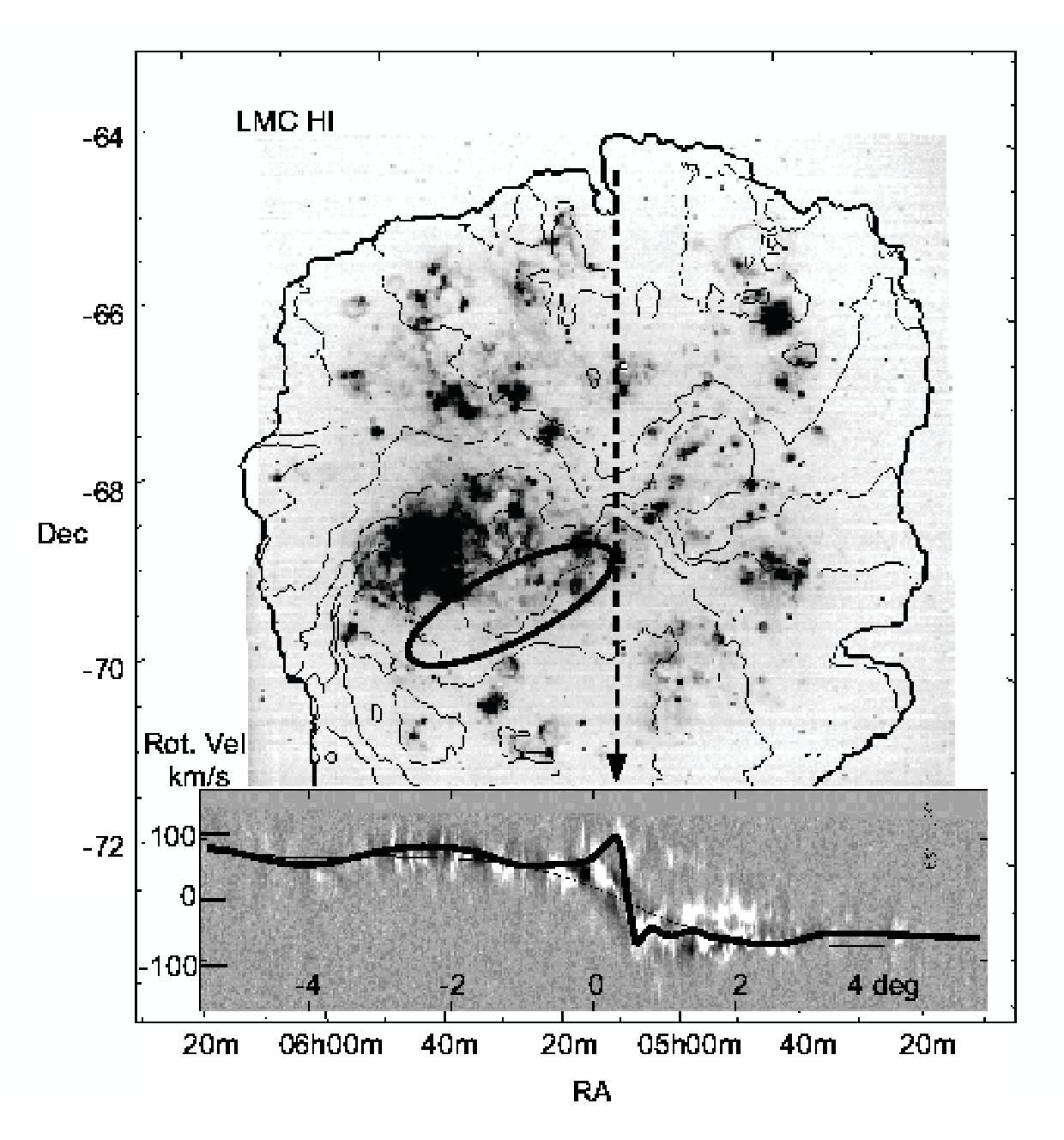}  
\caption{The HI velocity field of the LMC superposed on an \ha\ image,
and a position-velocity diagram across the kinematical major axis
(Kim et al. 1998: Courtesy of S. Kim).
The ellipse indicates the position of the optical bar.
The thick line in the PV diagram traces the rotation curve, corrected
for the inclination angle of 33\deg. }
\end{center}
\label{Fig6}
\end{figure} 

\subsection{Irregular Galaxies: Interacting and Merging}

Rotation curves for irregular galaxies are not straightforward.
Some irregular galaxies exhibit quite normal rotation curves, such as
observed for a ring galaxy NGC 660, amorphous galaxy NGC 4631
and NGC 4945 (Sofue et al. 1999a).

The interacting galaxy NGC 5194 (M51) shows a very peculiar rotation
curve, which declines more rapidly than  Keplerian at
$R\sim 8 - 12$ kpc.
This may be due to inclination varying with the radius, e.g. warping.
Because the galaxy is viewed nearly face-on ($i = 20^\circ$), a slight warp
causes a large error in deriving the rotation velocity.
If the galaxy's outer disk at 12 kpc has an inclination as small as
$i \sim 10^\circ$, such an apparently steep velocity decrease would be observed
even for a flat rotation.

When galaxies gravitationally interact, they tidally distort each other, and
produce the pathological specimens that had until recently defied
classification.  In an innovative paper, Toomre (1977; see also
Toomre \& Toomre 1972; Holmberg 1941)
arranged eleven known distorted galaxies $``$in rough order of completeness of
the imagined mergers" starting with the Antennae (NGC 4038/39) and ending with
NGC 7252.  Observers rapidly took up this challenge, and Schweizer (1982)
showed that NGC 7252 is a late-stage merger, in which the central gas disks of
the two original spirals still have separate identities.

There is now an extensive literature both observational and computational
(Schweizer 1998 and references therein; Barnes \& Hernquist 1992, 1996;
Hibbard et al.  2000) that make it possible to put limits on the initial
masses, the gas quantities, the time since the initial encounter, and the
evolutionary history of the merger remnant. Equally remarkable, tidal tails
can be used as probes of dark matter halos (Dubinski et al.  1999).

Many nearby galaxies are also products of mergers, and hence have been
extensively studied.  NGC 5128 (Cen A) has a long history of velocity
observations, with few signs of being completely untangled yet (H$\alpha$ long
slit, Graham 1979; H$\alpha$ Fabry-Perot, Bland-Hawthorn et al. 1997; CO,
Phillips et al. 1987; HI, van Gorkom et al. 1990;  PN, Hui  et al. 1995).

The starburst dwarf galaxy NGC 3034 (M82)
shows an exceptionally peculiar rotation property (Burbidge et al. 1964;
Sofue et al. 1992).
It has a normal steep nuclear rise and rotation velocities  which have a
Keplerian decline beyond the nuclear peak. This may arise from a tidal
truncation of the disk and/or halo by an encounter with M81 (Sofue 1998).

The past and present history of the Milky Way and the Local Group is written
in the warp of the Milky Way (Garcia-Ruiz et al. 2000), in the tidal disruption
of the Sagittarius dwarf (Ibata et al. 1995), in the mean retrograde motion of
the younger globular clusters (Zinn 1993), in the tidal streams in the halo
(Lynden-Bell and Lynden-Bell, 1995), and in the orbits of the Magellanic
Clouds and the Magellanic stream (Murai \& Fujimoto 1980;
Gardiner \& Naguchi 1996).
For an excellent discussion $``$Interactions and Mergers in the Local
Group" with extensive references, see Schweizer (1998).

\subsection{Polar Ring Galaxies}

Polar ring galaxies provide an unique opportunity to probe the
rotation and mass distribution perpendicular to galaxy disks, and hence the
three dimensional distribution of the dark matter
(Schweizer et al. 1983; Combes \& Arnaboldi 1996; Sackett \& Sparke 1990, et al
1994; van Driel et al. 1995).
The conclusion of Schweizer et al. (1983) from emission line velocities,
that the halo mass is more nearly circular than flattened, has been
contested by Sackett \& Sparke (1990), based on both emission and absorption
line data. However, the data are limited and velocity uncertainties are large,
so the conclusions are not robust.

A major surprise  comes from the study of the polar ring galaxy NGC 4650A.
Arnaboldi et al. (1997; see also van Gorkom et al. 1987) discovered an
extended HI disk {\it coplanar with the ring}, which twists from almost
edge-on to more face-on at large radii. The K-band optical features and the
HI velocities can be fit simultaneously with a model in which spiral arms are
present in this polar disk. Hence the polar ring is a very massive disk.
This result strengthens previous suggestions that polar ring galaxies are
related to spirals (Arnaboldi et al. 1995; Combes \& Arnaboldi 1996).

\section{THE FUTURE}

Most of what we know about rotation curves we have learned in the last
fifty years, due principally to instrumental and computational advances.
It is likely that these advances will accelerate in the future. We can look
forward to an exciting future. Specifically,

 1.  Extinction-free rotation kinematics in the central regions will come
from high-$J$ CO line spectroscopy and imaging using ALMA, the Andes
large mm- and sub- mm wave interferometer at 5000 m altitude.
This array  will produce high spatial (0.01 arcsec) and
 high velocity ($\delta V<1$ \kms) resolution.

 2. Extinction-free measurements will also come from eight- and
ten-meter class telescopes using Br $\gamma$, H$_{2}$ molecular, and
other infrared lines in K-band and longer-wavelength spectroscopy.

3. VLBI spectroscopic imaging of maser sources will teach us more about
super massive black holes and rotation and mass distribution in nuclear disks.

 4. Rotation of the Galaxy, separately for the disk and bulge,
will be directly measured from proper motions, parallaxes, and hence,
distances, and radial velocities of maser sources using micro-arcsecond
radio interferometry. VERA (VLBI Experiment for Radio Astrometry) will
become a prototype.
This facility will also derive an accurate measurement of
the Galactic Center distance, $R_0$.

 5. Optical interferometry may permit us to $``$watch" a nearby
spiral (M31, M33, LMC, etc.)  rotate on the plane of the sky,
at least through a few microarcsecond.
Radio interferometry of maser stars will be used to
directly measure rotation on the sky of the nearest galaxies, from measures
of proper motion (hence, distances) and radial velocities.

 6. Rotation curves will be determined for galaxies at extremely high
redshift; with luck we will observe protogalactic rotation and dynamical
evolution  of primeval galaxies.
This may be a task for successive generations of space telescopes.

 7. We may be lucky and ultimately understand details of barred spiral
velocity fields from spectroscopic imaging.
We may be able to separate the disk, bulge and bar
potentials by fitting the number of parameters necessary for describing
bar's mass and dynamical properties.

 8. Polar ring kinematics will be understood, especially halo kinematics
perpendicular to the disk, and therefore, 3-D halo structure.
We are certain to
learn details of galaxies which are unexpected and hence surprising.
We may even be luckier and learn something new about the Universe.

 9. Sophisticated methods of analysis, perhaps involving line shapes and
velocity dispersions, will produce more accurate rotation curves for
large samples of spirals. These will lead to more tightly constrained
mass deconvolutions.
Distribution of dark and luminous matters within the halo,
disk, bulge, and core will be mapped in detail from more sophisticated
mass-to-luminosity ratio analyses.

 10. Dark halos will finally be understood.   We will know their extent, and
their relation to the intracluster dark mass. We may even know the
rotation velocity of the halo. Will the concept of a $``$rotation curve"
apply at such large distances from the disk? Will we learn if our
halo brushes the halo of M31?

 11. We will ultimately know what dark matter is,
the major constituent of the Universe when measured in mass.
Elementary particle physics will teach us its origin and physical properties.

 12. Perhaps we will be able to put to rest the last doubt about the
applicability of Newtonian gravitational theory on a cosmic scale, or
enthusiastically embrace its successor.

\vskip 10mm

Acknowledgements:
The authors thank Dr. Yoichi Takeda for assisting in gathering and
selecting the references from a huge number of related papers in the decades.
They thank Drs. Linda Dressel, Jeffrey Kenney, Stacy McGaugh,
and Rob Swaters for references and helpful comments.
They also thank Mareki Honma for references, Jin Koda and Kotaro Kohno
for NMA data analyses.

\vskip 10mm
This is an unedited draft of a chapter submitted for publication in the
Annual Review of Astronomy and Astrophysics, Volume 39, 2001
\parindent=0pt
\parskip=0pt

(http://astro.annualreviews.org/).

\newpage

\def\r{\hangindent=1pc \noindent}

\noindent{\bf References}
\parindent=0pt
\parskip=0pt

\r Aaronson M, Huchra J, Mould J 1980. {\it Ap. J.} 237:655

\r Aaronson M, Mould J. 1986. {\it Ap. J.} 303:1

\r Ables JG, Forster JR, Manchester RN, Rayner PT, Whiteoak JB, Mathewson DS, Kalnajs AJ, Peters WL, Wehner H. 1987. {\it MNRAS}. 226:157

\r Amram P, Balkowski C, Boulesteix J, Cayatte V, Marcelin M, Sullivan III WT. 1996. {\it Astron. Astrophys}. 310:737

\r Amram P, Boulesteix J, Marcelin M, Balkowski C, Cayatte V, Sullivan III WT. 1995. {\it Astron. Astrophys. Supp.} 113:35

\r Amram P, Le Coarer E, Marcelin M, Balkowski C, Sullivan III WT, Cayatte V. 1992. {\it Astron. Astrophys. Supp.} 94:175

\r Amram P, Marcelin M, Balkowski C, Cayatte V, Sullivan III WT, Le Coarer E. 1994. {\it Astron. Astrophys. Supp.} 103:5 year?

\r Amram P, Sullivan III WT, Balkowski C, Marcelin M, Cayatte V. 1993. {\it Ap. J. Lett.} 403:59

\r Argyle E. 1965. {\it Ap. J.} 141:750

\r Arnabaldi M, Oosterloo T, Combes F, Freeman KC, Koribalski B. 1997. {\it Astron. J.} 113:585

\r Arnaboldi M, Freeman KC, Gerhard O, Matthias M, Kudritzki RP, Mendez RH, Capaccioli M, Ford H. 1998. {\it Ap. J.} 507:759

\r Arnaboldi M, Freeman KC, Sackett PD, Sparke LS, Capaccioli M. 1995. {\it Planetary \& Sp. Sci.} 43:1377

\r Ashman KM. 1992. {\it PASP.} {\it Ap. J.} 104:1109

\r Athanassoula E Bureau 1999. {\it Ap. J.} 522:699

\r Athanassoula E, Bosma A, Papaioannou S. 1987. {\it Astron. Astrophys. Supp.} 179:23

\r Babcock HW. 1939. {\it Lick Obs. Bull.} 19:41

\r Baldwin JE, Lynden-Bell D,    Sancisi R. 1980. {\it MNRAS}. 193:313

\r Bally J, Stark AA, Wilson RW,  Henkel C. 1987. {\it Ap. J. Suppl.} 65:13

\r Barnes JE, Hernquist L. 1992. {\it Annu. Rev. Astron. Astrophys}. 30:705

\r Barnes JE, Hernquist L. 1996. {\it Ap. J.} 471:115

\r Barteldrees A, \& Dettmar R-J. 1994.  {\it Astron. Astrophys. Suppl.} 103: 475

\r Barton EJ, Bromley BC, Geller MJ. 1999. {\it Ap. J. Lett.}   511:25

\r Begeman KG, Broeils AH, Sanders RH. 1991. {\it MNRAS}. 249:523

\r Begeman KG. 1989. {\it Astron. Astrophys}. 223:47

\r Bender R, Burstein D, Faber SM. 1992. {\it Ap. J.} 411:153.

\r Bender R. 1990. {\it Astron. Astrophys}. 229:441

\r Bershady MA, Haynes MP, Giovanelli R, Andersen DR. 1999.  in {\it Galaxy Dynamics}. eds. DR Merritt, JA Sellwood, M Valluri. 182:263. San Francisco: Astron. Soc. Pacific

\r Bershady MA. 1997. in {\it Dark Matter in the Universe}. eds. M Persic, P Salucci. 117:537. San Francisco: Astron. Soc. Pacific

\r Bertola F, Bettoni D, Buson LM, Zeillinger WW. 1990. In {\it Dynamics and Interactions of Galaxies}. eds. ER Weilen. 249. Springer: Heideberg

\r Bertola F, Bettoni D, Zeillinger WW. 1992. {\it Ap. J. Lett.}  401:79

\r Bertola F, Cappellari M, Funes JG, Corsini EM, Pizzella A, Vega Bertran JC. 1998. {\it Ap. J. Lett.} 509:93

\r Bertola F, Cinzano P. Corsini EM, Pizzella A, Persic M, Salucci P. 1996. {\it Ap. J. Lett.} 458:L67

\r Bertola F, Galletta G.1978. {\it Ap. J. Lett.} 226:115

\r Binney J, Gerhard OE, Stark AA, Bally J, Uchida KI 1991. {\it MNRAS.} 252:210

\r Binney J, Merrifield M. 1998. {\it Galactic Astronomy}. Princeton Univ. Press

\r Binney J, Tremaine S. 1987. {\it Galactic Dynamics}. Princeton Univ. Press

\r Binney J. 1982. {\it Annu. Rev. Astron. Astrophys.} 20:399

\r Blais-Ouellette S, Amram P, Carignan C. 2000b. {\it Astron. J.} submitted.

\r Blais-Ouellette S, Carignan C, Amram P, Cote S. 2000a. {\it Astron. J.} in press

\r Bland-Hawthorn J, Freeman KC, Quinn PJ. 1997. {\it Ap. J.} 490:143

\r Blitz L. 1979. {\it Ap. J.} 227:152

\r Bosma A, van der Hulst JM, Athanassoula E, 1988. {\it Astron. Astrophys} 198:100

\r Bosma A. 1981a. {\it Astron. J.} 86:1825  

\r Bosma A. 1981b. {\it Astron. J.} 86:1791

\r Bosma A. 1996. in {\it Barred Galaxies}. eds. R Buta, DA Crocker, BG Elmegreen. {\it PASP Conf. Series.} 91:132.

\r Brand J, Blitz L. 1993. {\it Astron. Astrophys}. 275:67.

\r Braun R, Walterbos RAM, Kennicutt JR, Tacconi LJ. 1994. {\it Ap. J.} 420:558

\r Brinks E, Skillman ED, Terlevich RJ, Terlevich E. 1997. {\it ApSS}. 248:23

\r Broeils AH. 1992. {\it Astron. Astrophys}. 256:19

\r Burbidge EM,  Burbidge G. R. 1960. {\it Ap. J.} 132:30

\r Burbidge EM, Burbidge GR, Crampin DJ, Rubin VC, Prendergast KH. 1964. {\it Ap. J.} 139:1058

\r Burbidge EM, Burbidge GR. 1975. in {\it Stars and Stellar Systems IX: Galaxies and the Universe}. eds. A Sandage, M Sandage, J Kristian. University of Chicago Press, p. 81

\r Bureau M, Athanassoula E. 1999. {\it Ap. J.} 522:686

\r Burstein D, Bender R, Faber SM, Nolthenius R. 1997. {\it Astron. J.}  114:4

\r Burstein D, Rubin VC, Ford Jr WK, Whitmore BC. 1986. {\it Ap. J. Lett.}   305:11

\r Burton WB, Gordon MA. 1978. {\it Astron. Astrophys.} 63:7

\r Buta R, Crocker  DA, Elmegreen BG. 1996. eds. {\it Barred Galaxies}.  PASJ Conf. Series. 91

\r Buta R, Purcell GB, Cobb ML, Crocker DA, Rautiainen P, Salo H. 1999. {\it Astron. J.} 117:778

\r Buta R, van Driel W, Braine J, Combes F, Wakamatsu K, Sofue Y, Tomita A. 1995 {\it Ap. J.} 450:593

\r Cayatte V, van Gorkom JM, Balkowski C, Kotanyi C. 1990. {\it Astron. J.} 100:604

\r Carignan C, Beaulieu S. 1989. {\it Ap. J.} 347:760

\r Carignan C, Freeman KC. 1985. {\it Ap. J.} 294:494

\r Carignan C, Puche D. 1990. {\it Astron. J.}  100:394

\r Carignan C, Puche D. 1990. {\it Astron. J.}  100:641

\r Carignan C. 1985. {\it Ap. J.} 299:59

\r Carter D, Jenkins CR. 1993. {\it MNRAS}. 263:1049

\r Casertano S, van Gorkom JH. 1991. {\it Astron. J.}  101:1231

\r Casertano S. 1983. {\it MNRAS}. 203:735

\r Clemens, DP. 1985. {\it Ap. J.} 295:422

\r Combes F, 1992. {Annu. Rev. Astron. Astrophys.} 29:195

\r Combes F, Arnaboldi M. 1996. {\it Astron. Astrophys}. 305:763

\r Combes F, Mamon GA, Charmandaris V. 2000. eds. {\it Dynamics of Galaxies: from the Early Universe to the Present (XVth IAP Meeting: PASP. Conf. Series)}. Vol. 197

\r Corradi RLM,  Boulesteix J, Bosma A, Amram P, Capaccioli M. 1991. {\it Astron. Astrophys.} 244:27

\r Courteau, S. 1997. {\it Astron. J.} 114:2402

\r Courteau, S, Rix, HW. 1999 {\it Ap. J.}  513:561

\r de Blok E, McGaugh, SS, Rubin VC. 2001. in {\it Galaxy Disks and Disk Galaxies}.  eds. J Funes, E Corsini.  {\it PASP. Conf. Series}, in press 

\r de Blok WJG, McGaugh SS, van der Hulst JM. 1996. {\it MNRAS} 283:18

\r de Blok WJG, McGaugh SS. 1998. {\it Ap. J.} 508:132

\r de Vaucouleurs G,   Freeman KC. 1973. {\it Vistas in Astronnomy.} 14:163

\r de Vaucouleurs G. 1959. in {\it Handbuch der Physics, Astrophysik IV} 53:310

\r de Zeeuw T, Franx M. 1991. {\it Annu, Rev. Astron. Astrophys.} 29:239

\r Deguchi S, Fujii T, Izumiura H, Kameya O, Nakada Y, Nakashima J, Ootsubo T. Ukita N. 2000. {\it Ap. J. Suppl.} 128:571 

\r Dressler A. 1984. {\it Annu. Rev. Astron. Astrophys.} 22:185

\r Dubinski J, Mihos JC,  Hernquist L. 1999. { Ap. J.} 526:607

\r Einasto J, Saar E, Kaasik A, Chernin AD. 1974. {\it Nature}. 252:111

\r Evans NW, Collett JL. 1994. {\it Ap. J. Lett.}   420:67

\r Faber SM, Gallagher JS. 1979. {\it Annu. Rev. Astron. Astrophys.} 17:135

\r Ferrarese L, Ford HC. 1999. {\it Ap. J.} 515:583

\r Fich M, Blitz L, Stark A. 1989. {\it Ap. J.} 342:272

\r Fischer P. et al. 2000. {\it Astron. J.} submitted

\r Fisher D. 1997. {\it Astron. J.} 113:950

\r Forbes, DA. 1992. {\it Astron. Astrophys. Supp.} 92:583

\r Franx M, Illingworth GD, de Zeeuw T. 1991. {\it Ap. J.} 383:112

\r Franx M, Illingworth GD. 1988. {\it Ap. J. Lett.} 327:55

\r Fraternali F,  Oosterloo T, Sancisi R, van Moorsel G. 2000. in {\it Gas and Galaxy Evolution}, VLA 20th Anniversary Conference, J.E. Hibbard, M.P. Rupen \& J.H. van Gorkom (eds.), in press

\r Funes SJ, Corsini EM. 2001.  eds.  {\it Galaxy Disks and Disk Galaxies (ASP. Conf. Series)} in press

\r Galletta G. 1987. {\it Ap. J.} 318:531

\r Galletta G. 1996.  in {\it Barred Galaxies}. eds. R Buta, DA Crocker, Elmegreen BG. {\it PASP. Conf. Series.} 91:429

\r Garcia-Burillo S, Guelin M,  Cernicharo J. 1993. {\it Astron. Astrophys.} 274:123

\r Garcia-Burillo S, Sempere MJ, Bettoni D. 1998. {\it Ap. J.}  502:235

\r Garcia-Ruiz I, Kuijken K,  Dubinski J. 2000. {\it MNRAS}. submitted

\r Gardiner LT, Noguchi M. 1996. {\it MNRAS.} 278:191

\r Genzel R, Eckart A, Ott T, Eisenhauer F. 1997. {\it MNRAS}. 291:219

\r Genzel R, Pichon C, Eckart A, Gerhard OE, Ott T. 2000. {\it MNRAS.} 317:348

\r Gerhard OE. 1993. {\it MNRAS}. 265:213

\r Gerssen J. 2000. thesis Rijksuniversiteit Groningen

\r Ghez A, Morris M, Klein BL, Becklin EE. 1998. {\it Ap. J.} 509:678.

\r Gilmore G, King IR, van der Kruit PC. 1990. in {\it The MIlky Way as a Galaxy}.   University Science Books: Mill Valley, CA

\r Graham JA. 1979. {\it Ap. J.} 232:60

\r Greenhill LJ, Gwinn CR, Antonucci R, Barvainis R. 1996. {\it Ap. J. Lett.} 472:L21 

\r Greenhill LJ, Moran JM, Herrnstein JR. 1997. {\it Ap. J. Lett.} 481:L23 

\r Guhathakrta P, van Gorkom JH, Kotanyi CG, Balkowski C. 1988. {\it Astron. J.}  96:851

\r H\'{e}raudeau Ph, Simien F. 1997. {\it Astron. Astrophys}. 326:897

\r Handa T, Nakai N, Sofue Y, Hayashi M, Fujimoto M, 1990. {\it Publ. Astron. Soc. Jaapn}. 42:1

\r Haschick AD, Baan WA, Schneps MH, Reid MJ, Moran JM,  Guesten R. 1990. {\it Ap. J.}, 356:149. 

\r Haynes MP, Hogg DE, Maddalena RJ, Roberts MS, van Zee L. 1998. {\it Astron. J.}  115:62

\r Heisler J, Merritt D, Schwarzschild M. 1982. {\it Ap. J.} 258:490

\r Heraudeau P, Simien F. 1997. {\it Astron. Astrophys}. 326:897

\r Hernquist L, Barnes JE. 1991. {\it Nature}. 354:10

\r Herrnstein JR, Moran JM, Greenhill LJ, Diamond PJ, Inoue M, Nakai N, Miyoshi M, Henkel C, Riess A.  1999. {\it Nature}. 400:539

\r Hibbard JE, Vacca WD, Yun MS. 2000. {\it Astron. J.} 119:1130

\r Hoekstra H. 2000. thesis Rijksuniversiteit Groningen

\r Holmberg E. 1941. {\it Ap. J.} 94:385

\r Honma M, Sofue Y, Arimoto N. 1995. {\it Astron. Astrophys.} 304:1

\r Honma M, Sofue Y. 1996. {\it Publ. Astron. Soc. Japan Lett.} 48:103. 

\r Honma M, Sofue Y. 1996. {\it Publ. Astron. Soc. Japan Lett.} 48:103 

\r Honma M, Sofue Y. 1997a. {\it Publ. Astron. Soc. Japan.} 49:453 

\r Honma M, Sofue Y. 1997b. {\it Publ. Astron. Soc. Japan.} 49:539 

\r Hui X, Ford HC,  Freeman KC, Dopita MA. 1995. {\it Ap. J.} 449:592

\r Hunter DA, Rubin VC, Gallagher III, JS. 1986. {\it Astron. J.} 91:1086

\r Hunter JH, Gottesman ST. 1996. in {\it Barred Galaxies}. eds. R Buta, DA Crocker, Elmegreen BG. {\it PASP. Conf. Series}. 91:398

\r Ibata, RA, Gilmore G, Irwin MJ. 1995. {\it MNRAS.} 277:781

\r Irwin JudithA,  Sofue Y.  1992. {\it Ap. J. Lett.} 396:L75.

\r Irwin JudithA, Seaquist ER. 1991. {\it Ap. J.} 371:111

\r Ishizuki S, Kawabe R, Ishiguro M, Okumura SK, Morita KI, Chikada Y, Kasuga T, Doi M. 1990. {\it Ap.J.} 355:436

\r Izumiura H, Deguchi S, Hashimoto O, Nakada Y, Onaka T, Ono T, Ukita N, Yamamura I. 1995. {\it Ap. J.} 453:837 

\r Izumiura H. Deguchi S, Fujii T, Kameya O, Matsumoto S, Nakada Y, Ootsubo T, Ukita N. 1999. {\it Ap. J. Suppl.} 125:257 

\r Jacoby GH, Ciardullo R, Ford HC. 1990. {\it Ap. J.} 356:332

\r Jedrrejewski R, Schechter P. 1989. {\it Astron. J.} 98:147

\r Jobin M, Carignan C. 1990. {\it Astron. J.}  100:648

\r Kamphuis J, Briggs F.  1992. {\it Astron. Astrophys.} 253:335

\r Kaneko N, Aoki K, Kosugi G, Ohtani H, Yoshida M, Toyama K, Satoh T, Sasaki M. 1997. {\it Astron. J.}  114:94

\r Kannappan SJ, Fabricant DG. 2000. {\it Astron. J.}. in press

\r Kelson DD, Illingworth GD, van Dokkum PG, Franx M. 2000a. {\it Ap. J.}
31:159

\r Kelson DD, Illingworth GD, van Dokkum PG, Franx M. 2000b. {\it Ap. J.}
531:184

\r Kenney  J,  Young  SJ 1988. {\it Astrophys. J. Suppl} 66:261  

\r Kenney JDP,  Wilson CD, Scoville NZ, Devereux NA, Young JS. 1992. {\it Ap. J. L.} 395:L79

\r Kenney JPD, Faundez. 2000.  in {\it Stars, Gas and Dust in Galaxies: Exploring the Links}.  {\it PASP. Conf. Series}. in press

\r Kent SM. 1986. {\it Astron. J.}  91:1301

\r Kent SM. 1987. {\it Astron. J.} 93:816.

\r Kent SM. 1991. {\it Ap. J.} 378:131

\r Kent SM. 1992. {\it Ap. J.} 387:181

\r Kim S, Stavely-Smith L, Dopita MA, Freeman KC. Sault RJ, Kesteven MJ, McConnell D. 1998. {\it Ap. J.} 503:674.

\r Koda  J, Sofue Y, Wada K. 2000a. {\it Ap. J.} 532:214

\r Koda J,  Sofue Y, Wada K. 2000b. {\it Ap. J. L.} 531:L17

\r Kohno K, Kawabe R, Villa-Vilaro B. 1999.. {\it Ap. J.} 511:157

\r Kohno K. 1998. PhD. Thesis, University of Tokyo

\r Koo D. C. 1999. in {\it Proceedings of XIX Moriond Meeting on Building Galaxies: from the Primordial Universe to the Present}. eds. F Hammer, TX Thuan, V Cayatte, B Guiderdoni J, Tran Thanh Van (Gif-sur-Yvette: Editions Frontieres), in press

\r Kormendy J,  Richstone D. 1995. {\it Ann. Rev. Astron. Astrophys.} 33:581

\r Kormendy J, Westpfahl DJ, 1989. {\it Ap. J.} 338:752

\r Kormendy J.  1983. {\it  Ap. J.} 275:529

\r Kormendy J. 2001. in {\it Galaxy Disks and Disk Galaxies}.  eds. J Funes, E Corsini. {\it PASP. Conf. Series}, in press

\r Krabbe A, Colina L, Thatte N, Kroker H. 1997. {\it Ap. J.} 476:98

\r Kuijken K, Merrifield MR. 1993. {\it MNRAS}. 264:712

\r Kuijken K. 1996. {\it MNRAS}. 283:543

\r Kuno N, Nishiyama K, Nakai N, Sorai K, Vila-Vilaro  B. 2000. {\it Publ.
Astron. Soc. japan}. 52:775

\r Kyazumov GA, 1984. {\it AZh}. 61:846

\r Lake G, Schommer RA, van Gorkom JH. 1990. {\it Astron. J.}  99:547

\r Larkin JE, Armus L, Knop RA, Matthews K, Soifer BT. 1995. {\it Ap. J.}   452:599

\r Lequeux J. 1983. {\it Astron. Astrophys}. 125:394

\r Lin DNC, Pringle JE 1987. {\it Ap. J. L.} 320:L87

\r Lindblad B. 1959. in {\it Handbuch der Physics, Astrophysik IV} 53:21

\r Lindqvist M, Habing HJ, Winnberg A. 1992a. {\it Astron. Astrophys.} 259:118

\r Lindqvist M, Winnberg A, Habing HJ, Matthews HE.  1992b. {\it Astron. Astropjhys. Suppl.} 92:43

\r Lynden-Bell D, Lynden-Bell RM.  1995. {\it MNRAS.} 275:429

\r Maciejewski W, \& Binney J. 2000. IAU Symposium No. 205. {\it Galaxies and their Constituents at the Highest Angular Resolution} in press

\r Mathewson DS, Ford VL, Buchhorn M. 1992. {\it Ap. J. Suppl.} 81:413

\r Mathewson DS, Ford VL. 1996. {\it Ap. J. Supp.} 107:97

\r Mayall NU. 1951. in {\it The Structure of the Galaxy}.  Ann Arbor: University of Michigan Press p. 19

\r McGaugh SS, \& de Blok WJG. 1998. {\it Ap. J.} 499:66

\r Mediavilla E, Arribas S, Garcia-Lorenzo B, Del Burgo C. 1998. {\it Ap. J.}   488:682

\r Merrifield MR, Kuijken K. 1994. {\it Ap. J.} 432:75

\r Merrifield MR. 1992. {\it Astron. J.} 103:1552.

\r Merritt DR, Sellwood JA, Valluri M. 1999. eds. {\it Galaxy Dynamics}. San Francisco: Astron. Soc. Pacific

\r Mestel L. 1963.  {\it MNRAS}. 126:553

\r Milgrom M, Braun E. 1988. {\it Ap. J.} 334:130

\r Milgrom M. 1983. {\it Ap. J.} 270:371

\r Miyoshi M, Moran J, Herrnstein J, Greenhill L, Nakai N, Diamond P, Inoue M. 1995. {\it Nature}. 373:127.

\r Mo HJ, Mao S, White SDM. 1998. {\it MNRAS}. 295:319

\r Moellenhoff C, Matthias M, Gerhard OE. 1995. {\it Astron. Astrophys}. 301:359

\r Mulchaey JS, Regan MW. 1997. {\it Ap. J. Lett.} 482:L135

\r Murai T, Fujimoto M 1980. {\it Publ. Astron. Soc. Japan}. 32:581

\r Nakai N, Inoue M, Miyoshi M  1993. {\it Nature}. 361:45

\r Nakai N, Kuno N, Handa T, Sofue Y. 1994. {\it Publ. Astron. Soc. Jpn.} 46:527


\r Nelson AH. 1988. {\it MNRAS}. 233:115

\r Nishiyama K, Nakai N. 1998. IAU Symposium No. 184 {\it The central regions of the Galaxy and galaxies} p.245

\r Noguchi N. 1988. {\it Astron. Astrophys.} 203:259

\r Oestlin G, Amram P, Masegosa J, Bergvall N, Boulesteix J.  1999. {\it Astron. Astrophys. Supp.} 137:419

\r Oka T, Hasegawa T, Sato F, Tsuboi M, Miyazaki A, 1998. {\it Ap.J.S.} 118:455

\r Oort JH. 1940. {\it Ap. J.} 91:273

\r Ostriker JP,  Peebles PJE. 1973. {\it Ap. J.} 186:467

\r Ostriker, JP, Peebles, PJE, Yahil A. 1974. {\it Ap. J.} 193:L1O

\r Page T. 1952. {\it Ap. J.} 116:63 erratum 136:107

\r Pease FG. 1918. {\it Proc. Nat. Acad. Sci. US} 4:21

\r Peebles PJE. 1995. {\it Ap. J.} 449:52

\r Persic M, Salucci P, Stel F. 1996. {\it MNRAS}. 281:27

\r Persic M, Salucci P. 1988. {\it MNRAS}. 234:31

\r Persic M, Salucci P. 1990a. {\it MNRAS}. 247:349

\r Persic M, Salucci P. 1990b. {\it MNRAS}. 245:577

\r Persic M, Salucci P. 1995. {\it Ap. J. Supp.} 99:501

\r Persic M, Salucci P. 1997. eds. {\it Dark Matter in the Universe}. San Francisco: Astron. Soc. Pacific

\r Peterson CJ, Rubin VC, Ford Jr WK, Thonnard N. 1978.  {\it Ap. J.} 219:31

\r Phillips TG, Ellison BN, Keene JB, Leighton RB, Howard RJ, Masson CR, Sanders DB, Veidt B, Young K. 1987. { Ap. J. Lett.} 322:L73

\r Plummer HC. 1911. {\it MNRAS}. 71:460

\r Prada, F, Gutierrez CM, Peletier RF, McKeith CD. 1996. {\it Ap. J.} 463:9

\r Puche D, Carignan C, Bosma A. 1990. {\it Astron. J.}  100:1468

\r Puche D, Carignan C, Wainscoat RJ. 1991a. {\it Astron. J.}  101:447

\r Puche D, Carignan C, van Gorkom JH. 1991b. {\it Astron. J.}  101:456

\r Regan MW, Vogel SN. 1994. {\it Ap. J.} 43:536

\r Richstone D, Bender R, Bower G, Dressler A, Faber S, et al. 1998. {\it Nature} 395A:14 .

\r Richter O-G,  Sancisi R. 1994. {\it Astron. Astrophys}. 290:19

\r Rix H-W, Carollo CM, Freeman K. 1999. {\it Ap. J. Lett.} 513:L25

\r Rix H-W, Franx M, Fisher D, Illingworth G. 1992. {\it Ap. J. Lett.} 400:5

\r Rix H-W, Kennicutt RC, Braun R,  Walterbos RAM. 1995. {\it Ap. J.} 438:155

\r Rix H-W, White S. 1992. {\it MNRAS}. 254:389

\r Rix H-W, de Zeeuw PT, Cretton N, van der Marel RP, Carollo CM. 1997. {\it Ap. J.} 488:702

\r Roberts MA,  Rots AH. 1973. {\it Astron. Astrophys.} 26:483.

\r Roelfsema PR, Allen RJ. 1985. {\it Astron. Astrophys}. 146:213

\r Rogstad DH, Shostak GS. 1972. {\it Ap. J.} 176:315


\r Roscoe DF. 1999. {\it Astron. Asrophys.} 343:788

\r Rubin VC, Burbidge EM, Burbidge GR, Prendergast KH. 1965. {\it Ap. J.} 141:885.

\r Rubin VC, Burstein D, Ford Jr WK,  Thonnard N. 1985. {\it Ap. J.} 289:81

\r Rubin VC, Ford Jr WK, Thonnard N. 1980. {\it Ap. J.} 238:471

\r Rubin VC, Ford Jr WK, Thonnard N. 1982. {\it Ap. J.} 261:439

\r Rubin VC, Ford Jr WK. 1970. {\it Ap. J.} 159:379

\r Rubin VC, Ford Jr WK. 1983. {\it Ap. J.} 271, 556R

\r Rubin VC, Graham JA, Kenney JD. 1992. {\it Ap. J. Lett.} 394:9

\r Rubin VC, Graham JA. 1987. {\it Ap. J. Lett}. 316:67

\r Rubin VC, Haltiwanger, J. 2001. in {\it Galaxy Disks and Disk Galaxies}.  eds. J Funes, E Corsini.  {\it PASP. Conf. Series}, in press

\r Rubin VC, Hunter DA, Ford Jr WK. 1991. {\it Ap. J. Supp.} 76:153

\r Rubin VC, Hunter DA. 1991. {\it Ap. J. Supp.} 76:153

\r Rubin VC, Kenney JD, Boss AP, Ford Jr WK. 1989. {\it Astron. J.}  98:1246

\r Rubin VC, Kenny JDP, Young, JS. 1997. {\it Astron. J.}  113:1250

\r Rubin VC, Thonnard N, Ford Jr WK. 1982. {\it Astron. J.}  87:477

\r Rubin VC, Thonnard N, Ford WKJr.  1978. {\it Ap. J. Lett.} 225:L107

\r Rubin VC, Waterman AH,  Kenney JD, 1999. {\it Astron. J.}  118:236

\r Rubin VC, Whitmore B C, Ford Jr W K, 1988. {\it Ap. J.} 333:522

\r Rubin VC. 1994a. {\it Astron. J.} 107:173

\r Rubin VC. 1994b. {\it Astron. J.} 108:456

\r Rubin VC. 1987. in {\it Observational Evidence of Activity in Galaxies}. eds. EYe Khachikian, KJ Fricke, \& J Melnick. I.A.U. Symposium 121:461. Reidel:Dordrecht

\r Sackett PD, Rix H-W, Jarvis BJ, Freeman KC. 1994. {\it Ap. J.} 436:629

\r Sackett PD, Sparke LS, 1990. {\it Ap. J.} 361:408

\r Sackett PD, Sparke LS. 1994.  or et al

\r Sakamoto K, Okumura SK, Ishizuki S, Scoville NZ 1999. {\it Ap.J. S.} 124:403

\r Salo H, Rautiainen P, Buta R, Purcell GB, Cobb ML, Crocker DA, Laurikainen E. 1999. {\it Astron. J.}  117:792.

\r Salucci P, Ashman KM, Persic M. 1991. {\it Ap. J.} 379:89

\r Salucci P, Frenk CS. 1989. {\it MNRAS}. 237:247

\r Sancisi R, Allen RJ, Sullivan III WT. 1979. {\it Astron. Astrophys}. 78:217

\r Sancisi R. 2001. in {\it Galaxy Disks and Disk Galaxies}.  eds. J Funes, E Corsini.  {\it PASP. Conf. Series}, in press

\r Sanders RH, Verheijen MAW. 1998. {\it Ap. J.} 503:97

\r Sanders, RH. 1996. {\it Ap. J.} 473:117

\r Sargent AI,  Welch WJ. 1993. {\it Annu. Rev. Astron. Astrophys.} 31:297.

\r Sargent WLW, Schechter PL, Boksenberg A, Shortridge K. 1977. {\it Ap. J.} 212:326

\r Sawada-Satoh S, Inoue M, Shibata KM, Kameno S, Migenes V, Nakai N, Diamond PJ. 2000. {\it Publ. Astron. Soc. Jpn.} 52:421

\r Schaap WE, Sancisi R, Swaters RA 2000. {\it Astron. Astropys. L.} 356:49

\r Schinnerer E, Eckart A, Tacconi LJ, Genzel R, Downes D. 2000. {\it Ap. J.} 533:850. 

\r Schombert JM, Bothun GD, Schneider SE, McGaugh SS. 1992.  {\it Astron. J.} 103:1107

\r Schombert JM. Bothun GD. 1988. {\it Astron. J.} 95:1389

\r Schwarzschild M. 1954. {\it Astron. J.} 59:273

\r Schweizer F, Whitmore BC, Rubin VC.  1983. {\it Astron. J.} 88:909

\r Schweizer F. 1982. {\it Ap. J.} 252:455

\r Schweizer F. 1998. in {\it Galaxies: Interactions and Induced Star Formation},  Swiss Society for Astrop. Astron. XIV, eds. R. C. Kennicutt, Jr. F. Schweizer, J. E. Barnes, D. Friedli, L. Martinet, \& D. Pfenniger. Springer-Verlag Berlin

\r Scoville NZ, Thakker D, Carlstrom JE, Sargent AE 1993. {\it Ap. J. L} 404:L63

\r Shlosman I, Begelman MC, Frank, J  1990. {\it Nature}. 345:679

\r Sjouwerman LO, van Langevelde HJ, Winnberg A, Habing HJ.  1998. {\it Astron. Astrophys. Suppl.} 128:35

\r Sil'chenko KO. 1996. {\it Ast. L.} 22:108

\r Simard L, Prichet CJ. 1998. {\it Ap. J.} 505:96

\r Simkin SM. 1974. {\it Astron. Astrophys}. 31:129

\r Slipher VM. 1914. {\it Lowell Obs. Bull.} 62

\r Sofue Y, Honma M, Arimoto N. 1995. {\it Astron. Astrophys.} 296:33

\r Sofue Y, Irwin JudithA. 1992. {\it Publ. Astron. Soc. Jpn.} 44:353

\r Sofue Y, Koda J, Kohno K, Okumura SK, Honma M,  Kawamura A, Irwin JudithA. 2000.  {\it Ap. J. Lett.} in press.

\r Sofue Y, Reuter H-P, Krause M, Wielebinski R, Nakai N. 1992. {\it Ap. J.}   395:126

\r Sofue Y, Tomita A, Honma M, Tutui Y, Takeda Y. 1998. {\it Publ. Astron. Soc. Japan.} 50:427

\r Sofue Y, Tomita A, Honma M, Tutui Y. 1999b. {\it Publ. Astron. Soc. Jpn.} 51:737

\r Sofue Y, Tutui Y, Honma M, Tomita A, Takamiya T, Koda J, Takeda Y. 1999a.  {\it Ap. J.} 523:136

\r Sofue Y, Tutui Y, Honma M, Tomita A. 1997. {\it Astron. J.} 114:2428

\r Sofue Y. 1995. {\it Publ. Astron. Soc. Jpn.} 47:527

\r Sofue Y. 1996. {\it Ap. J.}. 458:120

\r Sofue Y. 1997. {\it Publ. Astron. Soc. Japan.} 49:17

\r Sofue Y. 1998. {\it Publ. Astron. Soc. Japan.} 50:227

\r Sofue Y. 1999. {\it Publ. Astron. Soc. Japan.} 51:445

\r Sorai K, Nakai N, Kuno N, Nishiyama K, Hasegawa T. 2000. {\it Publ.
Astron. Soc. Japan}. 52:785

\r Sorensen S-A, Matsuda T, Fujimoto M. 1976. {\it Astrophys. Sp. Sci.} 43:491

\r Sperandio M, Chincarini G, Rampazzo R, de Souza R. 1995. {\it Astron. Astrophys. Supp.} 110:279

\r Sridhar S,  Touma J. 1996. {\it MNRAS}. 279:1263

\r Steinmetz M, Navarro JF. 1999. {\it Ap. J.} 513:555

\r Swaters RA, Madore BF, Trewhella M. 2000. {\it Ap. J. Lett.} 356:L49.

\r Swaters RA, Sancisi R, van der Hulst JM. 1997. {\it Ap. J.} 491:140

\r Swaters RA, Schoenmakers RHM, Sancisi R, van Albada TS. 1999. {\it MNRAS}.    304:330

\r Swaters RA. 1999. thesis Rijksuniversiteit Groningen

\r Swaters RA. 2001. in {\it Galaxy Disks and Disk Galaxies}. eds. J Funes, E Corsini.  {\it PASP. Conf. Series}, in press

\r Takamiya T, Sofue Y. 2000. {\it Ap. J.} 534:670

\r Tecza M, Thatte N, Maiolino R. 2000. IAU Symposium No. 205. {\it Galaxies and their Constituents at the Highest Angular Resolution} in press

\r Thaker AR. Ryden BS. 1998. {\it Ap. J.} 506:93

\r Toomre A, Toomre J. 1972. {\it Ap. J.} 178:623

\r Toomre A. 1977. in {\it The Evolution of Galaxies and Stellar Populations}. eds. BM Tinsley, RB Larson. Yale University Observatory

\r Toomre A. 1982. {\it Ap. J.} 259:535

\r Tremaine S, Yu Q. 2000. {\it MNRAS}. submitted

\r Trimble V. 1987. {\it Annu. Rev. Astron. Astrophys.} 25:425

\r Trotter AS, Greenhill LJ, Moran JM, Reid MJ, Irwin JudithA, Lo K-Y. 1998. {\it Ap. J.} 495:740  

\r Tully RB,  Fisher JR. 1977. {\it Astron. Astrophys}. 54:661

\r Tully RB, Bottinelli L, Fisher JR, Goughenheim L, Sancisi R, van Woerden H.   1978. {\it Astron. Astrophys}. 63:37

\r Valluri M. 1993. {\it Ap. J.} 408:57

\r Valluri M. 1994. {\it Ap. J.}. 430:101

\r van Albada TS, Bahcall JN, Begeman K, Sancisi R. 1985. {\it Ap. J.} 295:305

\r van Albada TS, Kotanyi CG, Schwarzschild M. 1982 {\it MNRAS}. 198:303

\r van Driel W, Combes F, Casoli F, Gerin M, Nakai  N, Miyaji T, Hamabe M, Sofue Y, Ichikawa T, Yoshida S, Kobayashi Y, Geng F, Minezaki T, Arimoto N, Kodama T, Goudfrooij P, Mulder PS, Wakamatsu K, Yanagisawa K. 1995. {\it Astron. J.} 109:942

\r van Gorkom JH, Schechter PL, Kristian J.  1987. {\it Ap. J.} 314:457

\r van Gorkom JH, van der Hulst JM, Haschick AD, Tubbs AD. 1990. {\it Astron. J.}. 99:1781

\r van Moorsel GA. 1983. {\it Astron. Astrophys. Supp.} 54:1

\r van de Hulst HC, Raimond E, van Woerden H. 1957.  {\it Bull. Ast. Inst. Neth.} 14:1

\r van der Kruit PC, Allen RJ. 1978. {\it Annu. Rev. Astron. Astrophys.} 16:103

\r van der Kruit PC. 2001. in {\it Galaxy Disks and Disk Galaxies}.  eds. J Funes, E Corsini.  {\it PASP. Conf. Series}, in press

\r van der Marel M,  Franx M. 1993.  {\it Ap. J.} 407:525

\r van der Marel RP, Rix HW, Carter D, Franx M, White SDM, de  Zeeuw T. 1994.  {\it MNRAS}. 268:521

\r Vaughan JM. 1989. {\it The Fabry-Perot interferometer. History, theory, practice and applications.} The Adam Hilger Series on Optics and Optoelectronics, Bristol: Hilger.

\r Vega Baltran JC. 1999. Thesis Universidad de La Laguna

\r Verheijen MAW. 1997. thesis Rijksuniversiteit Groningen

\r Vogel SN, Rand RJ, Gruend RA, Teuben PJ. 1993. {\it PASP.} 105:666

\r Vogt NP, Forbes DA, Phillips AC, Gronwall C, Faber SM, Illingworth GD, Koo DC. 1996. {\it Ap. J. Lett.} 465:15

\r Vogt NP, Herter T, Haynes MP, Courteau S. 1993. {\it Ap. J. Lett.} 415:95V

\r Vogt NP. Phillips AC, Faber SM,  Gallego J, Gronwall C,  Guzman R, Illingworth GD, Koo DC, Lowenthal J D. et al. 1997. {\it Ap. J. Lett.} 479:121

\r Volders L. 1959. {\it BAN.} 1:323

\r Wada K, Habe A 1992. {\it MNRAS.} 258:82

\r Wada K, Habe A 1995. {\it MNRAS.} 277:433

\r Wada K, Sakamoto K, Minezaki T. 1998. {\it Ap.J.} 494:236

\r Walterbos RAM, Braun R, Kennicutt Jr RC. 1994. {\it Astron. J.} 107:184

\r Warner PJ. Wright MCH, Baldwin JE. 1973, {\it MNRAS}. 163:163

\r Watson WD, Wallim BK. 1994. {\it Ap. J.} 432:35

\r Weiner  B, Sellwood JA. 1999. {\it Ap. J.} 524:112

\r Weiner  BJ, Williams TB. 1996. {\it Astron. J.} 111:1156

\r Weiner BJ,  Williams TB,  van Gorkom JH,  Sellwood SA. 2000. {\it Ap. J.} accepted.

\r Weiner BJ, Sellwood SA, Williams  TB. 2000. {\it Ap. J.} accepted.

\r Westerlund BE. 1999. in {\it New Views of the Magellanic Clouds}, IAU Symposium No. 190, eds Y.-H. Chu, N. Suntzeff, J. Hesser, \& D. Bohlender. in press

\r Whitmore BC, Forbes DA, Rubin VC. 1988. {\it Ap. J.} 333:542

\r Whitmore BC, Forbes DA. 1989. {\it ApSS}. 156:175

\r Wilkinson MI, \& Evans NW. 1999. {\it MNRAS.} 310:645

\r Wolf M. 1914. {\it Vierteljahresschr Astron. Ges.} 49:162

\r Woods D, Madore BF, Fahlman GG. 1990. {\it Ap. J.} 353:90

\r Wozniak H, Pfenniger D. 1997. {\it Astron. Astrophys}. 317:14

\r Young  JS,  Xie  S,  Tacconi  L,  Knezek  P,  Vicuso  P, Tacconi-Garman  L,  Scoville  N,  Schneider  S,  et al.  1995. {\it Ap. J. Suppl} 98:219

\r Young JS, Scoville NZ. 1992. {\it Ann. Rev. Astron. Astrophys.} 29:581

\r Zaritsky D, Elston R, Hill JM. 1989. {\it Astron. J.} 97:97

\r Zaritsky D, Elston R, Hill JM. 1990. {\it Astron. J.} 99:1108

\r Zaritsky D, White SDM. 1994. {\it Ap. J.} 435:599

\r Zaritsky D. 1992. {\it Publ. Astron. Soc. Pac.} 104:831

\r Zinn R. 1993. in {\it  The globular clusters-galaxy connection.} ASP Conf. Series. eds. Graeme H. Smith, and Jean P. Brodie. 48:38

\end{document}